# Graphene oxide: new opportunities for optoelectronic, electronic and photonic chips

David J. Moss

Optical Sciences Centre, Swinburne University of Technology, Hawthorn, VIC 3122, Australia.

Email: dmoss@swin.edu.au

**Abstract**

As a derivative of graphene, graphene oxide (GO) was initially developed by chemists to emulate some of the key properties of graphene, but it was soon recognized as a unique material in its own right, addressing an application space that is not accessible to chemical vapor deposition based materials. Over the past decade, GO has emerged as a new frontier material with tremendous advances in its material fabrication and quality control over its properties. These in turn have led to rapid progress in GO based photonics, electronics, and optoelectronics concepts and devices, evoking new science and paving the way for many technological breakthroughs with exceptional performance. Here, we review the unique fundamental characteristics of GO, its thin film fabrication methods, patterning techniques, and mechanisms for tuning its material properties. This latter capability in particular has enabled novel advanced functional photonic, electronic, and optoelectronic devices. Understanding these insights is essential for designing and tailoring GO properties for state-of-the-art applications including solar energy harvesting, energy storage, medical diagnosis, image displays, and optical communications. We conclude by discussing the open challenges and exciting opportunities of this field, together with future prospects for major technological advancements and breakthroughs.

**Keywords:** Graphene oxide, 2D materials, integrated optics, nonlinear optics, photonic integrated circuits

# Introduction

Historically, the development of carbon materials has driven, and been accompanied by, many technological advances. Since the first discovery of buckminsterfullerene (so-called "bucky-balls") in 1985[1], followed by carbon nanotubes (CNTs) in 1991[2], low-dimensional forms of carbon materials have attracted tremendous interest. This trend intensified after the first experimental isolation of two-dimensional (2D) carbon monolayers – graphene – in 2004[3], and has expanded into many other forms of 2D carbon-based materials that contain more than just simple carbon atoms.

Graphene oxide (GO), as the most common derivative of graphene, consists of layered carbon networks decorated with oxygen-containing functional groups (OFGs). The history of GO dates back to 1859[4], and its investigation has experienced remarkable advancement in the past two decades (**BOX 1, panel a**), in tandem with a surge in interest in graphene-based materials in general. In the early days of graphene research, GO was a common route to making imperfect graphene[5]. Subsequently, the ability of GO to be dispersed in solution together with the interesting properties that arose when it was stacked into a lamellar structure, garnered significant attention in its own right. GO provides a flexible material platform for the attachment of a range of functional organic groups to the surface of a graphene-like carbon network, yielding functionalized graphene-based materials with a range of unique properties. In addition, facile solution-based synthesis processes have been developed for GO and its derivatives, which are attractive for the large-scale manufacturing of carbon-based materials and devices. Further, GO and its derivatives can be processed into diverse forms such as suspensions in solution, monolayer nanosheets, free-standing papers, thin membranes, and fillers in organic or inorganic nanocomposites (**BOX 1, panels b – f**). All of these have distinct morphological features that render them highly versatile for many high-tech applications.

================================================================================

## BOX 1 │ History and morphology of GO

The earliest known synthesis of GO was achieved by Oxford chemist B. Brodie, dating back to 1859[4] – over a century before the discovery of modern carbon materials such as fullerene and CNTs at the end of 20th century, followed by the rise of 2D materials such as graphene[3], transition metal dichalcogenides (TMDCs)[6], and black phosphorus (BP)[7] after 2004 (**panel a**). Brodie invented the method for oxidation of crystalline graphite and synthesized GO via exfoliation of graphite oxide. His method yielded many atomically thin GO sheets, which perhaps were the earliest synthesized 2D materials, although graphene itself was unknown at that time. In 1958, Hummers *et al.* made important modifications to the GO synthesis method, achieving a shorter reaction time, larger sheet sizes, and without the need for hazardous $ClO_2$ gas[8]. Nowadays, the Brodie and Hummers methods are still widely used for GO synthesis with only minor modifications. The ground-breaking discovery of graphene in 2004 revived the interest in GO, and increasingly, the evidence has shown that GO is in fact more than just a precursor of graphene, but rather has its own unique material properties that actually outperform graphene in many applications. The development of nanofabrication and characterization technologies for GO has led to many exciting advances in both scientific research and industrial uses in the past decade, such as observation of the atomic structure of GO on the nanoscale[9], the direct laser patterning of GO films[10], the synthesis of high-quality graphene via GO reduction[11], and the integration of 2D GO films on chip with precise control[12].

The bulk form of GO is graphite oxide, which is a brownish hydrophilic compound composed of layered GO sheets (**panel b**[13]). Unlike graphene, GO is soluble in aqueous and polar solvents, which allows for solution-based preparation and processing. **Panel c** shows the aqueous suspension of exfoliated GO sheets[14]. Depending on the concentration of GO sheets, the suspension's color can vary from light yellow to dark brown. **Panel d** shows an atomic force microscope (AFM) image of monolayer GO sheets having a thickness of about 1.2 nm[15]. By using different GO synthesis methods, the lateral size can be varied from several microns to several tens of microns. The exfoliated GO sheets can be used as a raw material to prepare free-standing papers (**panel e**[16]), thin membranes (**panel f**[17]), or composite materials[18] for different applications.

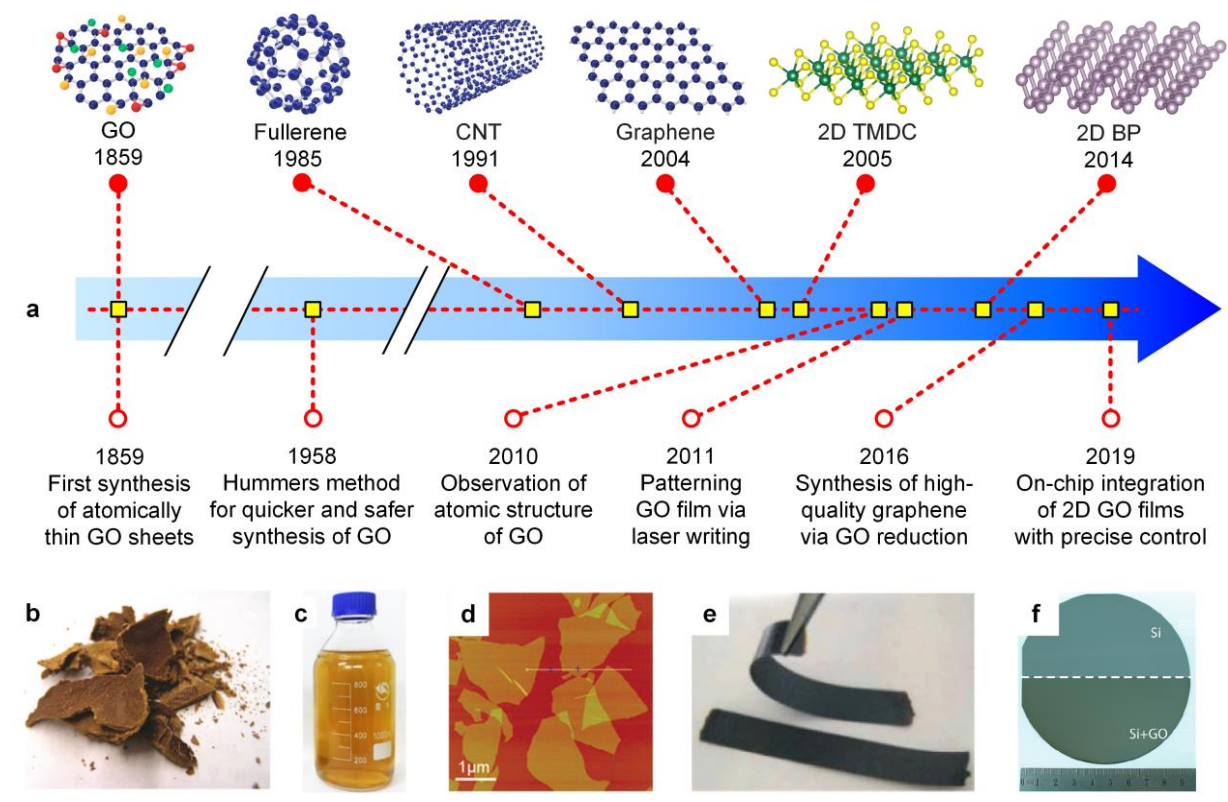

GO has a broad range of uses in many core industries that underpin our daily life. In the past decade, paralleling the growth of photonics, electronics, and their intersection – optoelectronics, GO devices have blossomed, spurred on by GO's material properties as well as the advancement of nanofabrication technologies. These devices have underpinned a wide range of technologies such as solar energy harvesting, energy storage, medical diagnosis, image displays, and optical communications. In this Review, we summarize the state-of-the-art of this field. Since both the electronic and optical properties of GO essentially reflect the response to electromagnetic waves at different frequencies, photonic and electronic applications of GO are closely related. While GO has been the subject of previous reviews[5,19-21], here we focus particularly on its applications to photonics, electronics, and optoelectronics, highlighting the significant progress and achievements over the past decade.

This review is structured as follows. We begin by comparing graphene, GO, and reduced GO (rGO), in terms of their atomic structure, optical bandgaps, and other fundamental properties. We then summarize the approaches for GO synthesis, device fabrication, and modifying its material properties. Next, we outline the optical, electronic, and optoelectronic properties of GO and rGO, and review the representative work in key areas. Finally, we discuss the current challenges and future prospects.

## Chemical structure & bandgap

In **FIG. 1**, we compare the atomic structure, bandgap, and high-resolution transmission electron microscopy (HRTEM) images of pristine graphene (PG), GO, and totally reduced GO (trGO). Since trGO has no residual oxygen-containing functional groups (OFGs), we examine it here in order to highlight the contrast between GO and reduced GO (rGO), together with PG compared with trGO.

As a derivative of PG, GO has different OFGs such as hydroxyl, carboxyl, and carbonyl groups (**FIG. 1a**) located either on the carbon basal plane or at the sheet edges. Compared with PG that consists of only $sp^2$-hybridized carbon atoms, some of the carbon atoms in GO are $sp^3$ hybridized through σ-bonding with the OFGs, leading to a heterogeneous structure. Depending on the different synthesis methods and preparation processes, there is a wide variability in the concentrations of $sp^2$ and $sp^3$ hybridizations, and this can also be affected by a reduction in OFGs or doping treatments. Together, these all enable a high degree of flexibility in changing GO's material properties.

Note that trGO has a defective carbon network that results from the complete reduction of GO that removes all the OFGs, and so its material properties are close to PG. The high yield of GO monolayers resulting from solution-based exfoliation provides an attractive method of mass-producing graphene-like materials. The difference in properties between trGO and PG mainly results from defects in the trGO. In the past decade, significant progress has been made towards obtaining high-quality graphene through the reduction of GO, led by methods based on wet chemistry[22] and microwave reduction[11].

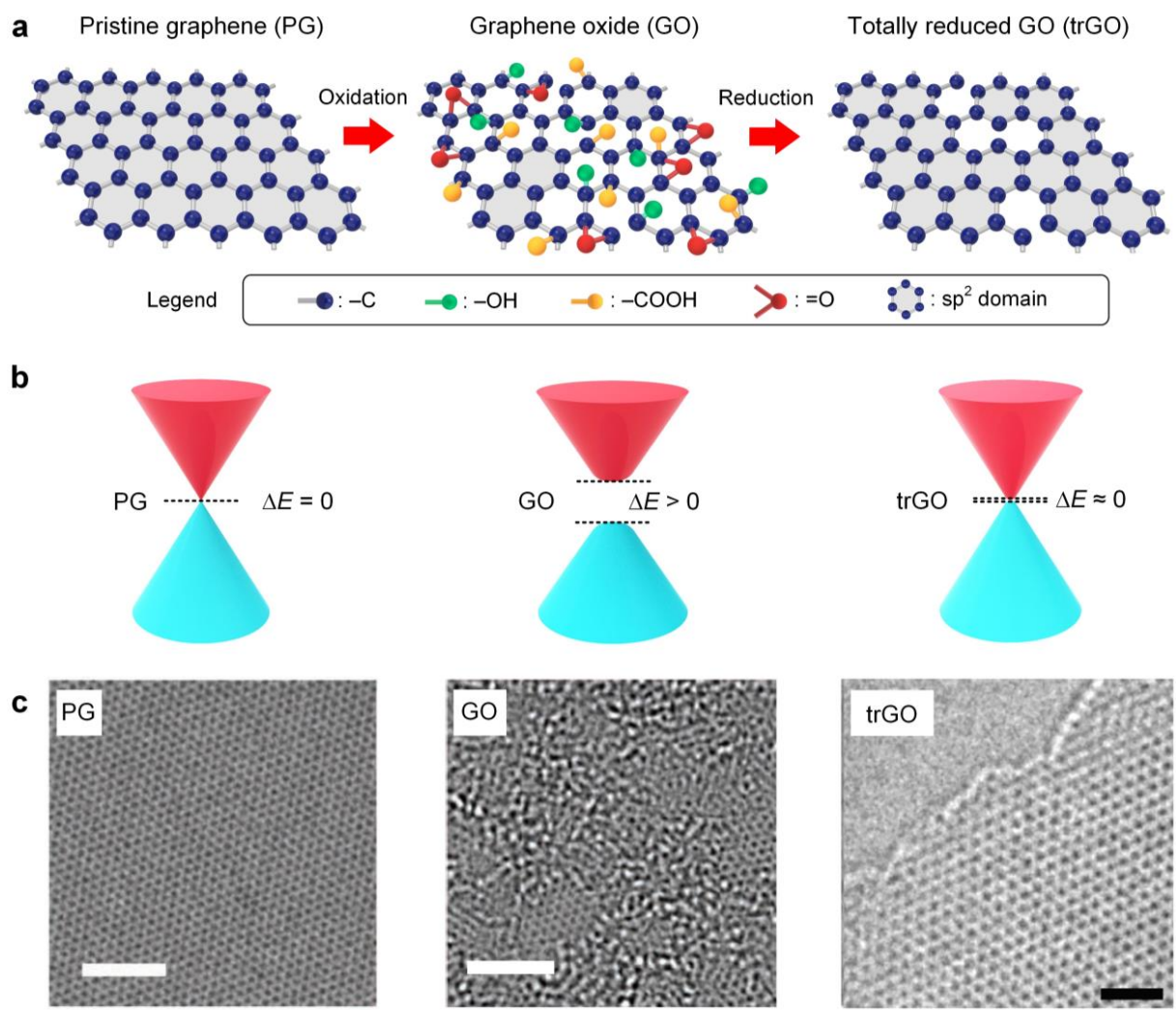

Fig. 1. **Comparison of pristine graphene (PG), graphene oxide (GO), and totally reduced GO (trGO).** **a|** Schematics of atomic structures. **b|** Schematics of bandgaps. **c|** High-resolution transmission electron microscopy (HRTEM) images. Scale bar for left and middle: 2 nm. Scale bar for right: 1 nm. Panel **c** left and middle adapted with permission from REF.[9], Wiley-VCH; panel **c** right adapted with permission from REF.[11], AAAS.

In contrast to PG that has a gapless Dirac cone, GO features an opened bandgap (**FIG. 1b**), arising from the isolated *sp²* clusters within the *sp³* C–O matrix, where the size of the *sp²* clusters determines its bandgap. Unreduced GO has a typical bandgap of 2 eV – 3.5 eV, which can be tuned by varying the *sp²* and *sp³* concentrations. This has been modeled by first principles theory[23] coupled with experimental characterization[17]. The reduction of GO can lead to an increase in the fraction of *sp²* bonds, thus reducing the bandgap. The ability to vary the bandgap of GO enables the tuning of its material properties including the refractive index, optical absorption, and electrical conductivity, all of which are key for photonic and electronic applications.

Due to the existence of OFGs, a monolayer GO film has a typical thickness of ~1 nm, which is much thicker than a monolayer of PG (~0.35 nm). Unlike PG that contains only a *sp²* carbon network with a highly uniform surface (**FIG. 1c**, **left panel**), GO has a hybrid carbon network that exhibits a highly disordered structure (**FIG. 1c**, **middle panel**). It contains many different regions including $sp^2$ regions having a preserved honeycomb structure, $sp^3$ regions with bonded OFGs, and defect regions, which all contribute to a relatively rough surface. Forming trGO through accurately controlled microwave reduction (**FIG. 1c**, **right panel**) results a highly ordered structure, almost the same as PG, reflecting thorough removal of the OFGs and nearly perfect reconstruction of the $sp^2$ carbon lattice after the reduction process.

## Material synthesis & device fabrication

### *GO film synthesis*

GO nanosheets can be produced on a large-scale by the chemical oxidization of graphite[24], where the hydrophobic graphene surface is converted to hydrophilic GO due to the attachment of OFGs. The hydrophilic property of GO allows for the dissolving of GO flakes in water, which enables various solution-based film fabrication methods as shown in **FIG. 2**. More importantly, it then becomes feasible to further modify the intrinsic material properties of GO by controlling the content of OFGs through various reduction methods[25]. These unique features make GO an excellent material platform for multifunctional devices.

A number of film synthesis methods have been demonstrated, which can be categorized according to their degree of control over the film thickness. We first review solution dropping methods that have coarse thickness control, such as drop casting (**FIG. 2a**), spray coating (**FIG. 2b**), and spin coating (**FIG. 2c**). Drop casting is the simplest method of fabricating GO films, where drops of GO solution are directly applied on a substrate and form a film after drying. The film thickness mainly depends on the concentration and volume of the GO solution, which are difficult to accurately control, and so result in a low uniformity. In contrast, smaller drop sizes can be realized by using a spray nozzle, with or without ultrasonification (**FIG. 2b**). The tiny droplets that result from this can significantly improve the uniformity and degree of control over the thickness by controlling the spray time. Alternatively, the thickness control and uniformity can also be improved by placing the substrate on a spinning disc and using a spin coating method (**FIG. 2c**), where the thickness can controlled by varying the GO solution concentration and spin speed. However, neither of these methods can create either continuous monolayers or freestanding films, due to their inability to accurately control the film thickness, together with the poor mechanical strength of the resulting GO films because of the random orientation of the GO flakes.

By comparison, the film uniformity and thickness can both be controlled better via filtration methods with controlled pressure (**FIG. 2d**). By varying the volume of GO flakes and controlling the GO solution concentration, it is possible to produce highly uniform GO films with controllable thicknesses. Since the GO films prepared in this way are formed on porous filter substrates, they typically need to be peeled off from the filter in order to transfer them to other substrates. It is also possible to obtain freestanding GO films for thick (> 200 nm) films to avoid any undesired influences of the substrate, such as for infrared applications[26]. On the other hand, filtration methods are limited in the size of the film areas they can produce. Further, they are relatively slow in the time needed to dry out the film, which inceases with film thickness. The Dr Blade method (**FIG. 2e**) can also be used to achieve a high degree of control over the thickness and uniformity of the GO film by controlling the gap between the blade and the substrate[27]. This method can produce large-area GO films, and is even amenable to roll-to-roll production methods. However, neither filtration nor the Dr Blade method are able to control the film accuracy down to the monolayer level.

A low-cost solution-based layer-by-layer GO film fabrication method has recently been developed[17]. In this method (**FIG. 2f**), a monolayer of positively charged polyelectrolyte polydiallyldimethylammonium chloride (PDDA) is coated on the negatively charged substrate (such as silica or silicon). Following this, a negatively charged GO monolayer is firmly attached to the positively charged PDDA layer, enabled by the electrostatic forces caused by the attachment of OFGs. By iterating the PDDA and GO deposition processes, multiple GO/PDDA layers can be conformally coated onto arbitrary substrates without the need for any transfer processes. This method is able to construct GO films with arbitrarily large areas and shapes. The substrates that have been used with this method include dielectrics (silicon wafers, glass slides, and cover slips), plastics (curved acrylic lenses and Polyethylene terephthalate (PET) films), metals (copper and silver), and even substrates composed of nanostructures (nanoholes in copper and silicon nanopillars)[28]. In addition, due to the strong electrostatic force present between the interlayers, the deposited GO films possess outstanding mechanical strength and excellent flexibility even after extensive bending and twisting tests. This layer-by-layer film deposition technique can yield large-area ultrathin GO films with a well controlled film thickness on a nanometer scale. This, together with the ability of conformal coating over diverse nanostructures, represents a significant advantage for large-scale manufacturable photonic, electronic, and optoelectronic devices.

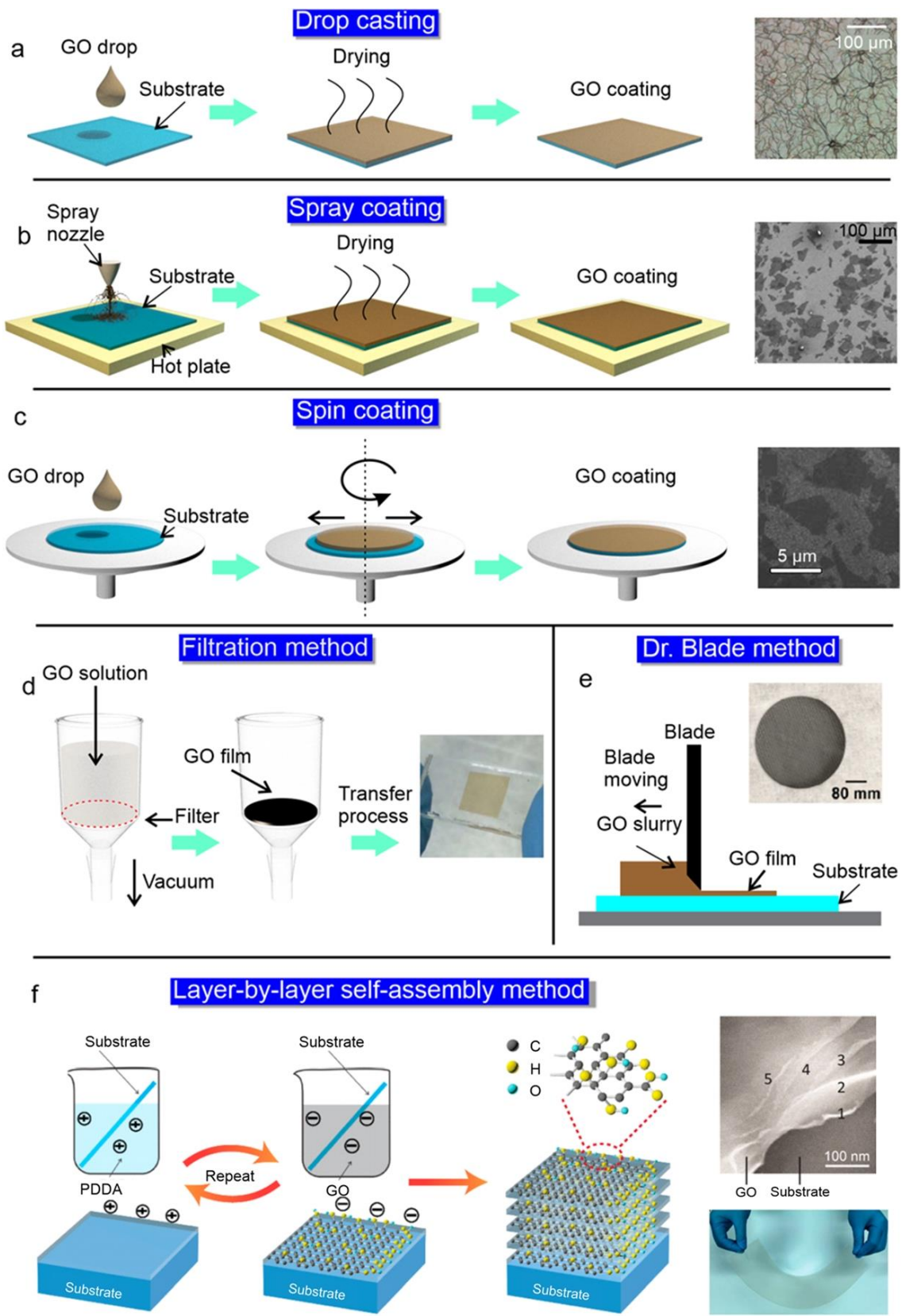

Fig. 2. **Fabrication methods for GO films. a|** Drop casting method, inset: transmission optical microscopy image of drop-casted GO film. **b|** Spray coating method, inset: scanning electron microscope (SEM) image of spray coated GO film. **c|** Spin coating method, inset: SEM image of spin coated GO film on cover glass. **d|** Filtration method, inset: photo of filtrated GO transferred on to a flexible substrate. **e|** Dr. Blade method, inset: photo of freestanding GO film prepared by using Dr.

Blade method. **f|** Layer-by-layer self-assembly method. Insets: SEM image of a 5 layer GO film on a silicon substrate (top) and GO coated on an A4 size flexible PET substrate (bottom). Panel **a** right adapted with permission from REF.[29], OSA publishing. Panel **b** right adapted with permission from REF.[30], IOP publishing. Panel **c** right adapted with permission from REF.[31], ACS publications. Panel **d** right adapted with permission from REF.[32], Springer Nature Limited. Panel **e** inset adapted with permission from REF.[27], ACS publications. Panel **f** adapted with permission from REF.[17], ACS publications.

## *GO reduction*

In order to convert GO to graphene-like materials, it is necessary to remove the OFGs, and this can be done through various reduction methods. The four reduction methods that are widely used include thermal reduction, chemical reduction, microwave reduction, and laser reduction[25].

Thermal reduction (**FIG. 3a**) is the most common, where the OFGs are removed by heating the GO films with a hot plate or oven at temperatures between 200 °C and 1000 °C[33]. At higher temperatures the carbon network begins to damage. In order to protect the carbon network from burning at high temperature, the thermal reduction can be performed in nitrogen ($N_2$), argon (Ar) or vacuum atmospheres. Although thermal reduction can process large volumes of GO, it cannot completely remove OFGs since C=O bonds are stronger than C=C bonds in the GO carbon lattice network. The degree of reduction can be increased by using chemical reduction in solution with reducing agents capable of targeting specific bonds (**FIG. 3b**). However, while chemical reactions can almost completely remove the OFGs[34,35], it is still necessary to fabricate rGO films after reduction, which is much more challenging than for GO films because of the hydrophobic properties of the rGO flakes.

More recently, microwave methods have been developed for GO reduction (**FIG. 3c**), and these typically consist of a two-step process[11]. First, the GO film is mildly pre-reduced using thermal reduction to improve its conductivity. This is followed by a complete reduction of the pre-reduced GO film in a microwave oven. This method has proven effective for removing the OFGs and restoring the carbon network, and can yield high quality rGO approaching that of PG. However, it is best applied to free-standing GO films since the electrical properties of the substrate can significantly affect the reduction process. Thus, it is not suitable for applications to semiconductor or metal substrates.

Lastly, laser photo-reduction (**FIG. 3d**) is currently regarded as one of the most effective and flexible methods of producing rGO[32,36], where the GO films are converted *in-situ* to rGO films. This method can be used to achieve film patterning, thus providing a useful approach to create functional devices[37,38 39-41]. In addition, laser treatment can simultaneously reduce and dope GO[42], and does not use harsh or toxic chemicals, nor does it require high temperatures.

During laser reduction, the OFGs are removed and $sp^2$ graphene domains are formed. To achieve high quality rGO, the OFGs need to be completely removed without destroying the $sp^2$ domains. Depending on the photon energy of the laser being used, laser photo-reduction can be classified into either photochemical or photothermal reduction[43,44]. It has been demonstrated that the photon energy threshold for photochemical reduction is 3.2 eV (i.e., λ = 390 nm), while a photon energy of < 3.2 eV (i.e., λ > 390 nm) generally yields photothermal reduction[43].

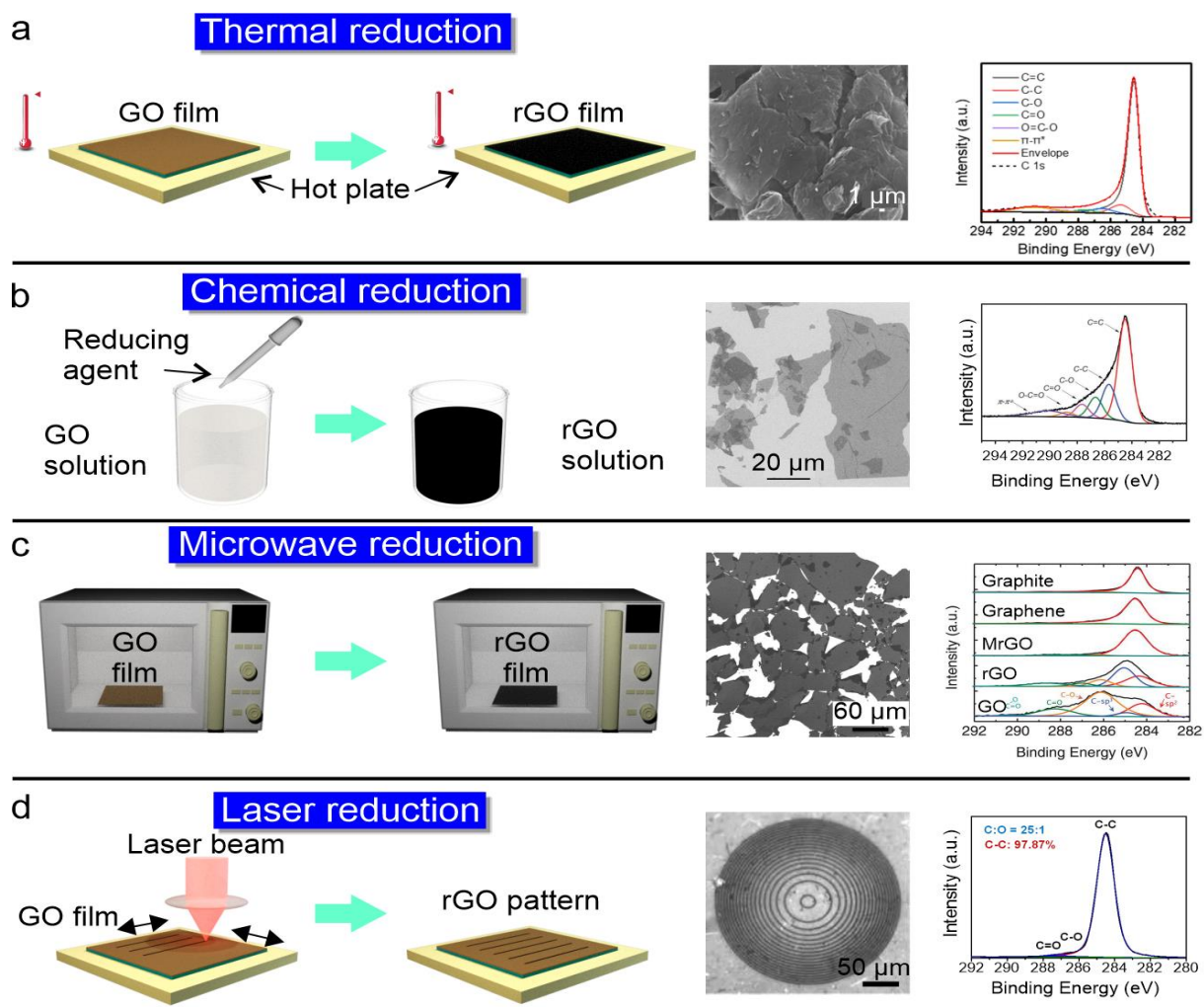

Fig. 3. **GO reduction methods. a|** Thermal reduction method, inset: SEM image of thermally reduced GO film and the X-ray photoelectron spectroscopy (XPS) spectrum. **b|** Chemical reduction method, inset: SEM image of chemically reduced GO film on cover glass and the XPS spectrum. **c|** Microwave reduction method, inset: SEM image of microwave reduced GO film and the XPS spectrum. **d|** Laser reduction method, inset: optical microscopy image of laser reduced GO film and the XPS spectrum. Panel **a** right adapted with permission from REF.[33], MDPI. Panel **b** right adapted with permission from REF.[34], RSC Publishing and REF.[35], MDPI. Panel **c** right adapted with permission from REF.[11], AAAS. Panel **d** right adapted with permission from REF.[36], OSA Publishing.

In the photothermal process, the light is absorbed by the entire GO structure, which heats up the GO molecules, increasing the temperature linearly with power. Several stages can be identified in the reduction process based on their temperature window[45]. Initially, water in the GO layers is evaporated, followed by the elimination of the OFGs at 200 °C, beginning with C-COOH groups due to their lowest binding energy. This is followed by the removal of C-OH groups at around 600 °C, after which the C-O-C groups are finally removed at 800 °C. The graphene $sp^2$ C=C network itself begins to break near 1000 °C. Increasing the temperature further results in the complete burning of the aromatic graphene network. Since the $sp^3$ C-C bond has a lower binding energy than any of the C-O bonds, breaking them introduces defects in the graphene network. Hence, photothermal reduction has a limited reduction ability with regards to C=O species, and it is challenging to achieve nearly complete reduction using this method alone without suffering from collateral damage to the graphene network.

In contrast, photochemical reduction can remove OFGs effectively and restore the aromatic domain in graphene through the non-resonant Stark effect[46]. Conventional photochemical reduction methods use either photocatalysts[47] or short wavelength light sources, and each have their trade-offs. Laser reduction without photocatalysts is advantageous, since the catalysts may affect the overall properties of the laser reduced GO. On the other hand, not using catalysts requires a short wavelength laser such as an excimer laser ($\lambda$ = 248 nm)[48] or even shorter extreme-ultraviolet sources ($\lambda$ = 46.9 nm)[49], where the penetration depth within the GO film is typically only a few 10's of nanometres. This makes it challenging to process very thick films.

In comparison, longer wavelength (visible and near infrared) femtosecond lasers have been applied to reduce GO films with minimal thermal effects and penetrate deeper into materials, achieving a relatively high C: O ratio (8:1)[50]. The high optical intensities involved with focused femtosecond laser irradiation can lead to multiphoton absorption, thus resulting in thermal relaxation that induces both photothermal and photochemical processes[43]. Multiphoton absorption can also lead to a higher spatial resolution, potentially reaching the nanometer scale[32], which is highly advantageous for fabricating photonic and electronic devices in a single step. During the femtosecond laser reduction, the material properties of GO, such as the refractive index, light absorption, conductivity, surface hydrophilicity, are modified by detaching the OFGs. This has been widely used for tailoring GO's material properties in many photonic and electronic devices[10,51].

## *Patterning and nanostructuring*

Patterning and nanostructuring of GO or rGO films are generally required during device fabrication to realize different functionalities[12,52,53], and are sometimes achieved in tandem with GO reduction processes. In **FIG. 4**, we summarize the methods for GO and rGO patterning and nanostructuring. Here nanostructuring is distinct from patterning in that it can be used to fabricate 3D nanostructures in contrast to planar patterns.

It is possible to create 2D patterns with relatively low resolution (in the range of 10's to 100's of microns) via the widely used inkjet printing method[54,55]. During this process (**FIG. 4a**), the substrate is placed on a hot plate, which helps to quickly dry out the printed ink droplets. The ink solution can be either GO or rGO solutions, each requiring different solvents that affect the viscosity and resolution of the inkjet printing process. If a GO solution is used and the temperature is set to below the reduction threshold (< 80 °C), it is possible to produce GO patterns. On the other hand, rGO patterns can be achieved by either directly using rGO solutions or by reducing the GO patterns. Since GO solutions can use water as a solvent, which has a very low viscosity, it is generally preferable to achieve high resolution patterns. Inkjet-printed GO/rGO patterns are generally used for sensing or energy storage applications[55,56], which don't require high spatial resolution. In comparison, as mentioned in the previously, focused laser beam photoreduction can produce micro/nanopatterns (**FIG. 4b**) to fabricate devices such as electrodes[57] and supercapacitors[58,59]. By increasing the laser power, the laser-reduced GO films can actually be ablated (**FIG. 4c**) to fabricate grooves or holes[51,60]. Patterning such grooves at designed positions can produce ultrathin gratings[51] and metalenses[60,61].

The layer-by-layer self-assembly method can produce conformally coated GO films on substrates that contain nanostructures (**FIG. 4d**), which is usually challenging for the methods based on layer transfer or physical vapor deposition. Along with the GO reduction methods, this allows for the further exploitation of the material properties of GO and rGO by enhancing the overlap with an optical mode or by taking advantage of resonant enhancement effects in nanocavities[28].

Conventional semiconductor patterning methods such as lithography and lift-off can also be used, either for selective film removal or for selective film coating over desired areas, as shown in **FIG. 4e** for a microring resonator (MRR) with a patterned GO film[12]. These methods often involve multiple steps, including coating, patterning, and developing the photoresist to create the pattern, followed by coating the GO films and finishing off by removing the photoresist. As compared with laser patterning methods, these approaches enable more accurate control of the size and position of the film patterns for coating small areas on a large substrate with either GO or rGO.

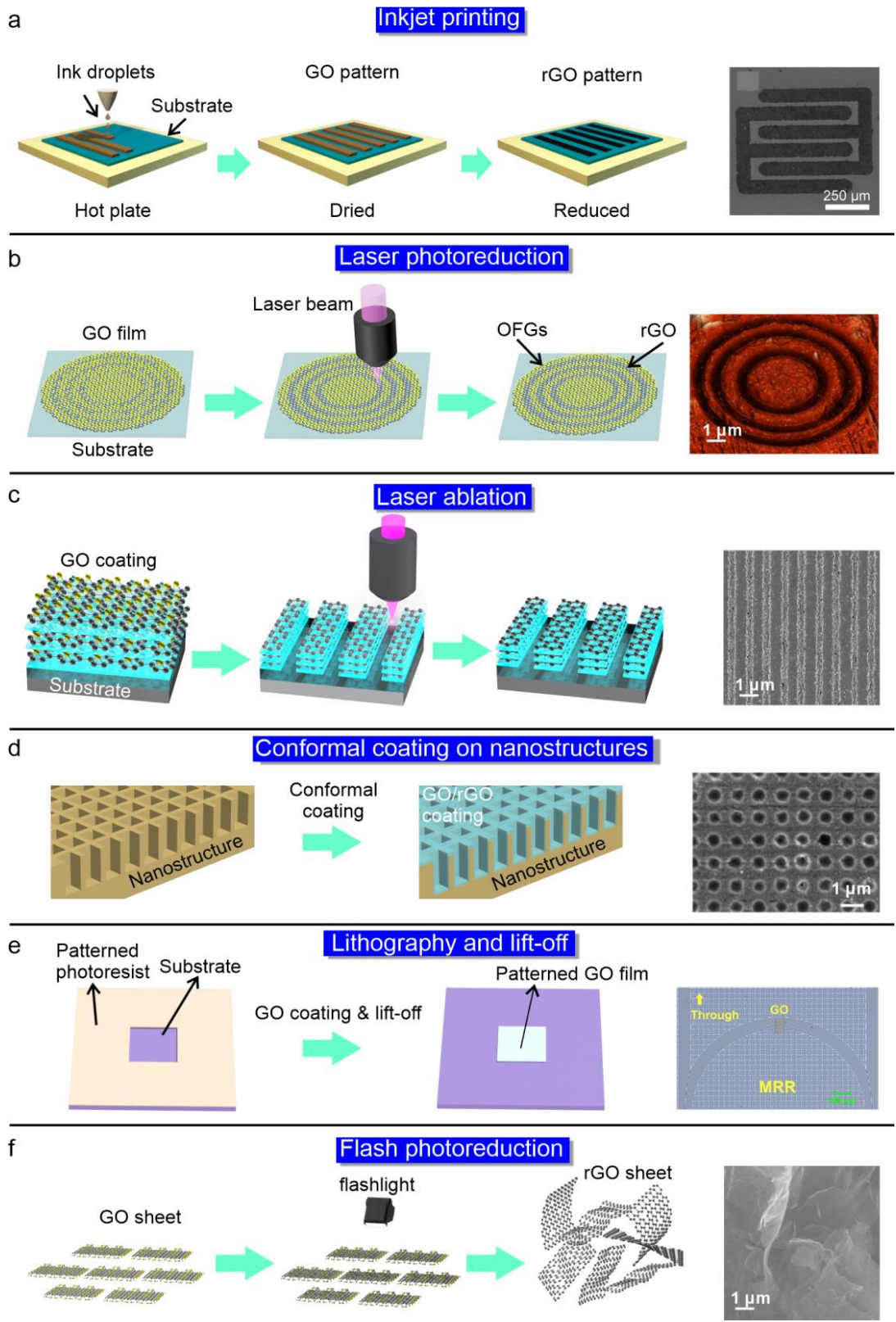

Fig. 4. **Patterning of GO/rGO film. a**| Production of micropatterns by using inject printing. Inset: SEM image of the inkjet printed electrode pattern. **b**| Generation of nanostructure by using laser photoreduction. Inset: AFM image of a graphene metalens. **c**| Generation of nanostructure by using laser ablation. Inset: SEM of a graphene nanograting. **d**| Conformal coating of GO metamaterials on prefabricated nanostructure. Inset: SEM image of conformally coated nanostructures. **e**| Patterning GO film on photonic devices using lithography and lift-off method. Inset: optical microscopy image of

GO film coated on a microring resonator (MRR). **f|** Generation of porous graphene material via flash photoreduction. Inset: SEM image of the porous graphene. Panel **a** right adapted with permission from REF.[54], Wiley-VCH. Panel **b** right adapted with permission from REF.[62], OEJ. Panel **c** right adapted with permission from REF.[51], Springer Nature Limited. Panel **d** right adapted with permission from REF.[28], Springer Nature Limited. Panel **e** right adapted with permission from REF.[12], Wiley-VCH. Panel **f** right adapted with permission from REF.[63], Elsevier Publishing.

Porous graphene nanostructures can be created efficiently and quickly by using a single flash light pulse[63] to reduce GO films (**FIG. 4f**). Flash lamp based photoreduction is similar, yet complementary, to laser photoreduction, offering similar advantages but with the added benefits such as improved efficiency, scalability, and accessibility. The pulsewidths of flash lamps (typically on the order of milliseconds) tend to be much longer than lasers and with much broader and incoherent spectral bandwidths. Hence, another advantage is that material ablation is much less likely than with femtosecond laser reduction, in situations where this is not desired.

In creating porous graphene nanostructures, the flashlamp pulse heats the GO film up to 2000 °C at an ultrafast rate up to $10^8$ K/min[63], which completely reduces the GO to a graphene-like material. During this process, gas products, including $CO_2$, CO, and $H_2O$, are rapidly released and expanded, resulting in the generation of nanopores inside the rGO films and 3D porous structures with tunable porosity. The porous structures, which can enhance light absorption by up to 99 %[64], are ideal for light trapping. Another advantage of this process is that the random distribution of pore sizes results in a very broadband absorption covering the entire solar spectrum. This can result in a near ideal black body absorption spectrum which is highly useful for applications such as solar energy harvesting[65]. Combining the large surface area of the porous structures with the high thermal conductivity of rGO allows solar water vapour generation[66]. Furthermore, the high electrical conductivity of these structures allows for the use of porous rGO as electrodes for electrochemical energy storage devices[27,63,67-69].

The patterning resolution of GO films is important in that it reflects the quality and level of fabrication, and is determined by both the patterning method and film quality. Inkjet printing has the lowest resolution on the order of microns. For laser patterning of GO films, the resolution approaches the diffraction-limit (i.e., 0.5λ / NA, where NA is the numerical aperture of focusing lens, and λ is the wavelength), which is typically > 200 nm. To achieve a high resolution, a high NA focusing lens combined with short working wavelengths is needed. For patterning of self-assembled GO films based on lithography followed by lift-off or alternatively on conformal coating of pre-patterned nanostructures, the patterning resolution is mainly limited by the minimum gap size at scales > 300 nm, and by the GO flake size at finer scales < 150 nm. The GO flake size, typically on the order of 100's nm, can be reduced further using modified oxidation and exfoliation methods[70], which can reduce the resolution limit yet further.

## Linear optical properties

The diverse range of OFGs present in GO results in a wide variability in its optical properties in comparison with PG. These properties can be tuned, or engineered, continuously by varying the degree of reduction, ultimately approaching that of PG. In this section, we summarize the linear optical properties of GO and rGO and review their relevant applications.

### *Solar absorption*

Like PG, rGO can absorb radiation over a very broad wavelength range from the ultraviolet to the terahertz regime[19,71]. The ability to achieve strong light absorbance in the solar spectral regime (**FIG. 5a**), along with a high carrier mobility, has enabled rGO to form the basis of high-performance solar devices. Although commercial silicon-based solar devices are mature, they are intrinsically limited in their light-energy conversion efficiency due to the poor light absorption of silicon across the solar spectrum. Another limitation of state-of-the-art solar devices is their need for special materials such as platinum and indium (e.g., for transparent indium tin oxide (ITO) electrodes), which are not only rare and expensive but also toxic. In contrast, rGO is derived from earth-abundant material and has high absorption across the entire solar spectrum.

A rGO solar absorber[51] is shown in **FIG. 5b**. Gratings with a thickness of 90 nm formed by rGO were fabricated by combining the layer-by-layer self-assembly method with direct laser writing (DLW). A high absorptivity of ~85% for unpolarized light across nearly the entire solar spectrum was achieved. The solar absorber also exhibited other advantages such as maintaining high light absorption over large incident angles and efficient heating under natural sunlight. A solar absorber based on a rGO film conformally coated on 3D metallic resonant structure was also reported[28]. The resonant structure was used to enhance the absorption specifically across the solar spectrum. This, together with the high thermal conductivity of rGO, resulted in a strong solar-selective absorption of omnidirectional light, yielding a solar-to-thermal conversion efficiency of up to ∼90.1%.

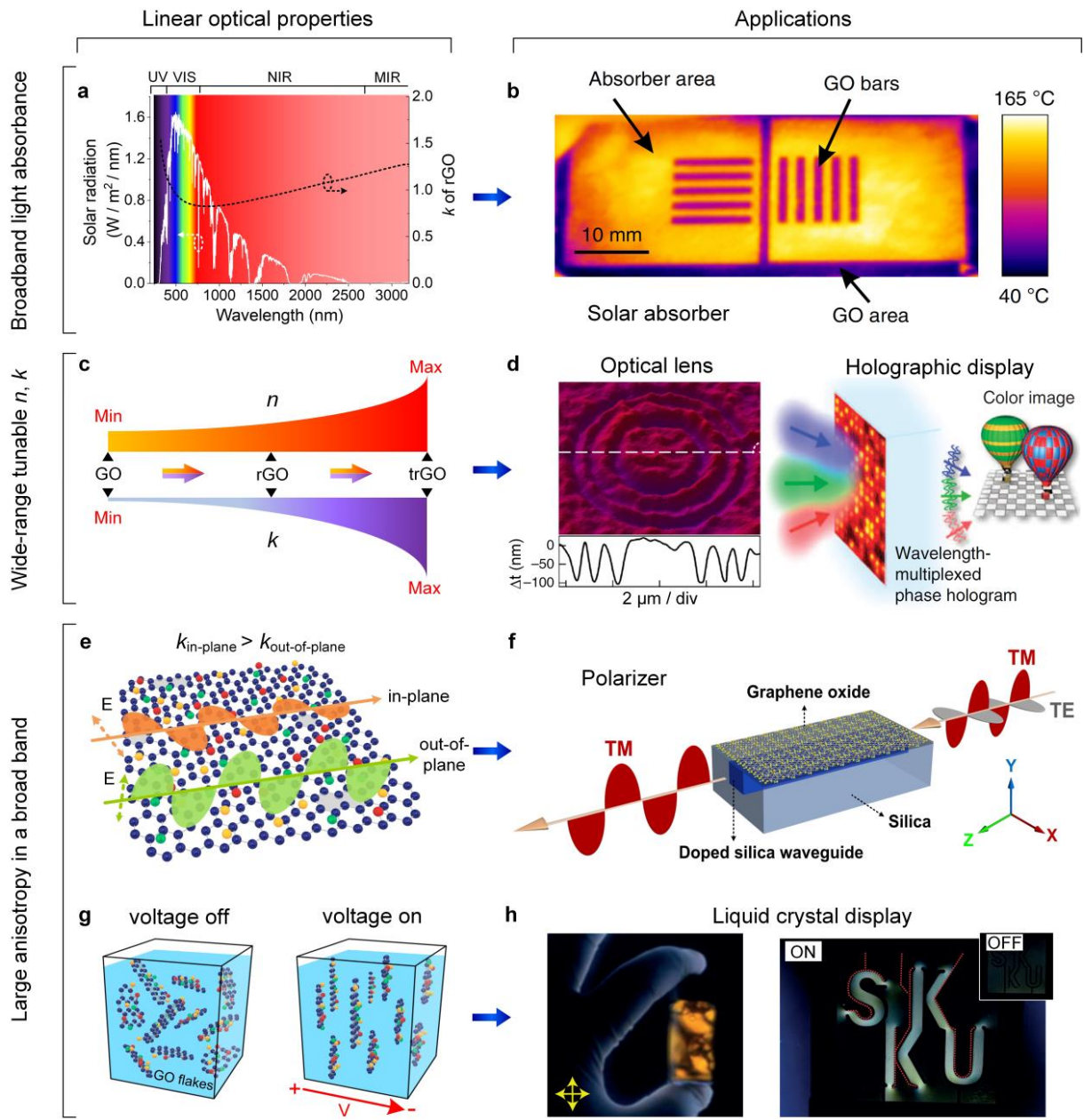

Fig. 5. **Linear optical properties and applications.** **a**| Solar radiation spectrum and broadband high material extinction coefficient $k$ of rGO. **b**| A rGO solar absorber. **c**| Change of refractive index $n$ and extinction coefficient $k$ from GO to trGO. **d**| A GO flat lens with 3D subwavelength focusing (left) and 3D holographic imaging based on photoreduction of GO (right). **e**| Schematic showing anisotropic loss of light propagating along a 2D GO film. **f**| A GO waveguide polarizer. **g**| Schematic showing electric-field induced orientation of GO liquid crystal (GO-LC). **h**| Birefringent textures of between crossed polarizers (left) and an electro-optic GO-LC switching (right). Panel **b** adapted with permission from REF.[51], Springer Nature Limited. Panel **d** left adapted with permission from REF.[32], Springer Nature Limited; panel **d** right adapted with permission from REF.[72], Springer Nature Limited. Panel **f** adapted with permission from REF.[12], Wiley-VCH. Panel **h** adapted with permission from REF.[73], Springer Nature Limited.

## *Tunable refractive index and extinction coefficient*

With a typical bandgap > 2 eV, pure unreduced GO displays an optical absorption that varies from the visible to the infrared wavelengths. In the near-infrared regime, a low material extinction coefficient $k$ of ~0.005 has been achieved for self-assembled GO films[12], which is over two orders of magnitude lower than PG. In the visible and near ultraviolet regimes, the $k$ of GO is typically > 0.1, mainly due to the π–π* transitions of C=C. On the other hand, GO displays a relatively moderate dispersion in its optical refractive index $n$, which is around 2 from the near-infrared to the mid-infrared regime[26]. For rGO, both $n$ and $k$ increase and approach that of PG as the degree of reduction increases (**FIG. 5c**). In contrast to bulk materials that usually have limited tuning ranges for $n$ and $k$, GO has a very large tuning range for both $n$ ($\Delta n \approx 0.8$, which is 1 – 2 orders of magnitude larger than conventional refractive materials[32]) and $k$ ($\Delta k \approx 2$). This allows wide-range phase and amplitude tuning in many photonic devices.

Optical lenses are fundamental to displays and imaging[32], and the ability to widely tune the $n$, $k$ of GO enables many novel forms of optical lenses with nanometer thicknesses and 3D subwavelength focusing capability[17,32,60-62]. These features are extremely challenging for conventional optical lenses that have a limited refractive index contrast. Moreover, GO lenses also show high stability in harsh environments[60] as well as flexibility in tuning the focal length[61]. With excellent, robust, and tunable focusing properties as well as facile and scalable fabrication, these devices are highly promising for practical applications.

An example of a thin GO flat lens[32] is shown in **FIG. 5d** (left). The flat lens, consisting of alternative sub-micrometer concentric rings formed by alternating GO and rGO regions, was patterned on a 200-nm-thick GO film via DLW. By virtue of the high index and absorption contrast between GO and rGO, both the phase and amplitude of the incident beam were effectively manipulated, allowing 3D subwavelength focusing ($\lambda^3/5$) below the diffraction limit. In addition, light focusing across a broad wavelength band from 400 nm to 1500 nm with an average focusing efficiency of > 32% was also achieved.

The ability to widely tune the $n$, $k$ of GO also offers new possibilities for holographic imaging devices in display systems. 3D holographic displays for wide-angle and full-color images (**FIG. 5d** (right)) were demonstrated based on multi-level refractive index modulation of rGO composites achieved with femtosecond optical pulses[72]. A wide viewing angle of 52 degrees was achieved for static 3D holographic imaging – an order of magnitude larger than CNT or metamaterial based holograms. Moreover, the spectrally flat variation in $n$, $k$ also allows for wavelength-multiplexed holograms for full-colour images. Recently, full color switchable optical displays were demonstrated based on an anisotropic layered pixel structure composed of alternating GO/rGO and PDDA layers on a silver-coated substrate[74]. Direct laser printing of all primary colors with a diffraction-limited pixel resolution (250 nm) in a large area (4 inch in diameter) was achieved by controlling the localized *n*, *k* and film thickness using laser photoreduction.

## *Optical anisotropy*

In its 2D form, GO displays an enormous anisotropy in its material properties, as also observed for 2D PG and TMDCs[75-77]. To date, most investigations of the anisotropy of 2D materials have focused on the linear material absorption and refractive index for applications such as polarization selection and liquid-crystal displays. The anisotropy of other properties such as the nonlinear optical Kerr coefficient, for example, remains an area of great interest.

Polarization control is fundamental to optical systems[78-80], and the huge absorption anisotropy and low dispersion of 2D materials are attractive for high performance polarization selective devices with high selectivity and broad bandwidths. As compared with in-plane polarized light, the absorption of out-of-plane polarized light in 2D GO films is much lower (**FIG. 5e**). This difference gradually decreases with increasing film thickness, as the 2D film evolves towards isotropic bulk-like material. Even for much thicker films (> 200 nm) than typical 2D materials, GO films can still maintain highly anisotropic light absorption for polarization-dependent devices[81,82]. In addition, this absorption anisotropy covers the visible to infrared regimes, enabling much more broadband polarization selective devices than bulk materials[12,26,81].

Waveguide polarizers based on patterned GO films integrated with doped silica waveguides[12] (**FIG. 5f**) achieved a broadband and high polarization dependent loss of up to ~53.8 dB across the visible and near infrared regimes. Layer-by-layer self-assembly coating methods enabled accurate film thickness control, which was combined with lithography and lift-off to yield control of the film size and location. As compared with state-of-the-art integrated silicon polarizers[80,83,84], GO waveguide polarizers not only have a much broader operation bandwidth but also feature simple designs with higher fabrication tolerance.

Many macroscopic uses of 2D materials require them to be dispersed and processed in the fluid phase, highlighted by liquid crystals (LCs)[85-87]. When there are a sufficiently large number of anisotropic particles suspended in a liquid medium, a dynamic LC transition can occur under external stimuli (**FIG. 5g**), forming the basis for LC displays. For PG with poor water solubility, the formation of LCs usually needs strong acids[88], in contrast to water-based LCs that are much easier to process. On the other hand, GO can be mass-produced from natural graphite via oxidation processes, and the hydrophilic OFGs in GO allow for monolayer exfoliation in water. These provide a viable route to realize carbon LCs in aqueous media[87].

An electro-optic switch based on GO LCs[73] is shown in **FIG. 5h**. Due to several limiting factors such as the electrophoretic drift, electrolysis of water, and GO reduction, it is usually difficult to control the alignment of nematic GO by applying electric fields. Song *et al.* overcame this by using low-concentration GO dispersions (0.05 vol%), where the interactions between the anisotropic GO flakes were weak enough to allow both microscopic ordering and macroscopic alignment of GO LCs under external electric fields. Based on this, electro-optic GO-LC switches with well-aligned, high-quality GO over large areas as well as with macroscopic sized electrodes were demonstrated.

## Nonlinear optical properties

The ever-increasing demand for data capacity and processing speed has motivated the use of nonlinear optical devices for optical communications[89,90], displays[91,92], astronomy[93,94], and quantum optics[95,96]. The ultrafast nonlinear optical response is highly useful for all-optical information processing since it avoids the need for slow and inefficient optical-electrical-optical (O-E-O) conversion, thus resulting in unparalleled processing speed[97,98].

Interest in the optical properties[99] of 2D materials – both linear and nonlinear – has grown significantly in the past decade. GO has unique advantages over other 2D materials, with its relatively large (up to 3.5 eV[100]) and direct material bandgap that enables broadband fluorescence for biosensing and molecular detection. It also has an ultrahigh Kerr optical nonlinearity (about 4 orders of magnitude larger than silicon) and relatively low linear absorption (over 100 times lower than PG in the infrared regime[101]), both of which are critical for nonlinear optical devices. The OFGs in GO can also break the material symmetry to produce a 2$^{nd}$ order optical nonlinearity, which is not observable in PG because of its centrosymmetry. In this section, we review the nonlinear optical properties of GO and rGO, together with their applications. Note that while fluorescence is typically not considered to be nonlinear, we discuss it here because it also involves a conversion of optical frequencies.

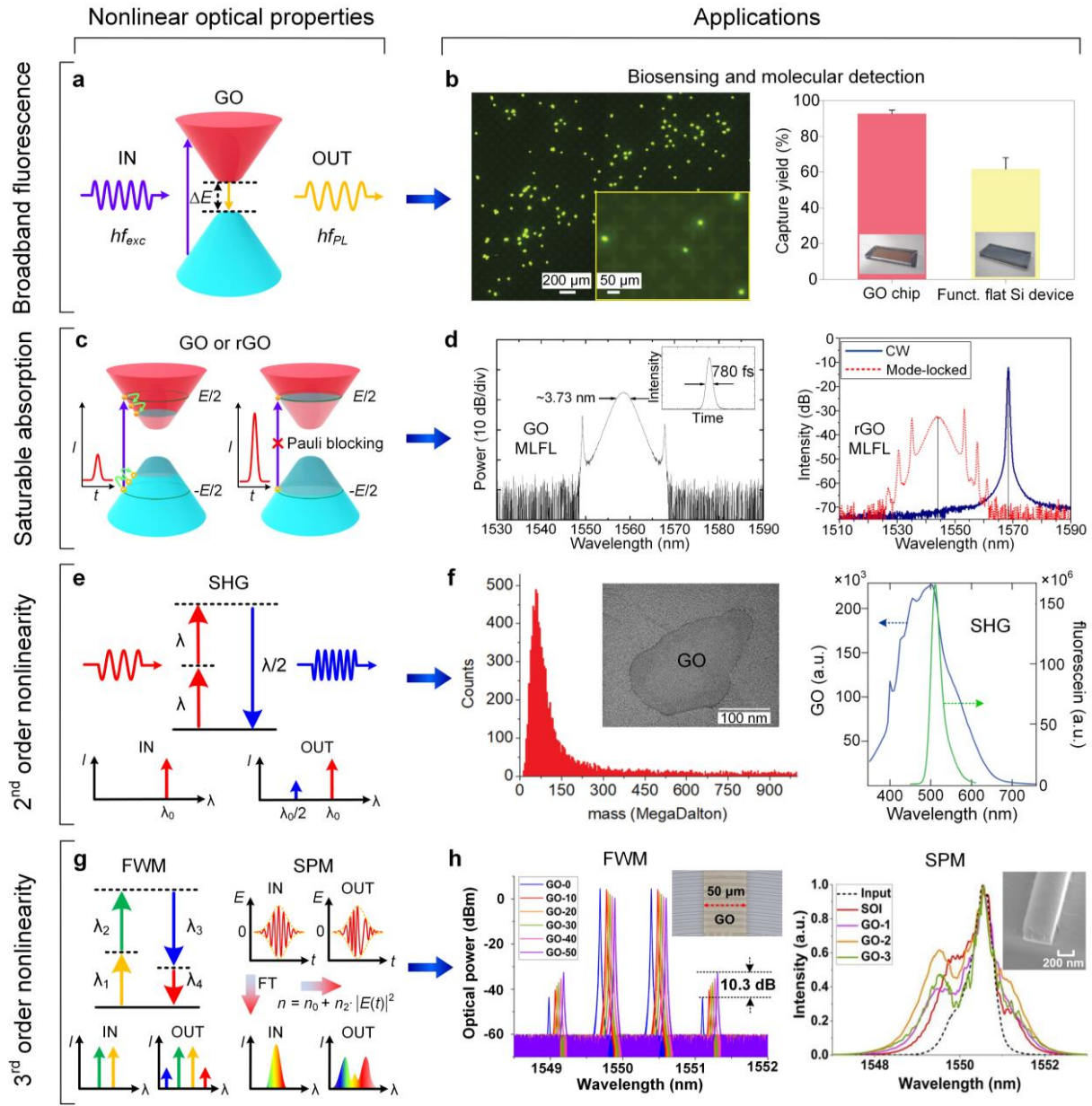

Fig. 6. **Nonlinear optical properties and applications. a|** Schematic band diagram for fluorescence process in GO. **b|** Fluorescence image of tumor cells captured by a GO chip (left) and comparison of capture yield between the GO chip and a functionalized flat silicon device (right). **c|** Schematic band diagram for the saturable absorption (SA) process in GO. **d|** Left: optical spectrum and temporal waveform of a mode-locked fiber laser (MLFL) using GO as a saturable absorber; Right: optical spectrum of a MLFL using rGO as a saturable absorber. **e|** Schematics showing energy and wavelength conversion of second harmonic generation (SHG). **f|** Left: mass histograms for GO samples, inset shows TEM image of the measured GO sample; Right: multiphoton excited fluorescence spectrum (blue) of GO with excitation at 800 nm, together with the two-photon-excited fluorescence spectrum (green). **g|** Schematics showing energy / wavelength conversion of four-wave mixing (FWM) process (left) and changes of optical pulses induced by self-phase modulation (SPM) (right). **h|** Left: enhanced FWM in a GO-coated integrated MRR; Right: enhanced SPM in a GO-coated silicon waveguide. Panel **b** adapted with permission from REF.[102], Springer Nature Limited. Panel **d** left adapted with permission from REF.[103], IOP Publishing. Panel **d** right adapted with permission from REF.[104], Elsevier Publishing. Panel **f** adapted with permission from REF.[105], Royal Society of Chemistry. Panel **h** left adapted with permission from REF.[53], Wiley-VCH; panel **h** right adapted with permission from REF.[106], ACS Publications.

## Fluorescence

Fluorescence, where a material absorbs photons with energies above its bandgap and subsequently radiates photons with energy near the bandgap after the excited photocarriers relax to lower energy states (**FIG. 6a**), is used for dyes, biosensing, gemology, medicine, mineralogy, printing, and many other areas[5,107]. In contrast to silicon's indirect bandgap and PG's zero bandgap, GO has a direct bandgap typically > 2 eV (**FIG. 1b**). This, together with the ability to tune the bandgap via reduction or doping, allows for tunable fluorescence across a broad band from the ultraviolet to the infrared regimes[5,20,108]. The broadband fluorescence of GO, along with its capability of direct wiring with molecules and high dissolubility in biomedia, has enabled sensitive biosensing and molecular detection[109].

GO has been widely used for fluorescent labeling. It has a fluorescence emission peak that is strongly dependent on the excitation wavelength when immersed in a polar solvent. This contrasts with organic dyes and inorganic quantum dots where the fluorescence emission peaks are independent of the excitation wavelength. The fluorescence wavelength in GO is affected by the giant red-edge effect that breaks Kasha's rule[110], making it highly promising for near-infrared biological imaging based on two-photon excitation spectroscopy. **FIG. 6b** shows an example of using GO for biomolecular detection[102], where a microfluidic chip with functionalized GO nanosheets on a patterned gold surface captured circulating tumor cells from blood samples of cancer patients. A high sensitivity of 73 ± 32.4% at 3–5 cells per ml blood was achieved, together with an improved capture yield over comparable functionalized flat silicon devices.

Although fluorescent, GO can also efficiently quench fluorescence via either charge transfer or the Förster resonance energy transfer (FRET) effect[107]. Interestingly, since GO has both fluorescence and quenching abilities, it can serve as either a donor or acceptor during the FRET processes. As a heterogeneous lattice, including both FRET donor and acceptor molecules on a large planar surface, GO enables different types of biosensors for biomolecule detection and DNA analysis. For example, fluorescent GO was used as a FRET donor for the detection of DNA hybridization[111], whereas quenching GO was employed as a FRET acceptor for selective and rapid detection of multiple DNA targets[112].

## Saturable absorption

Saturable absorption (SA), where the light absorption of a material decreases with increasing light intensity, has been widely used for short-pulsed lasers in industry and scientific research[99,113]. Bulk SA materials such as organic dyes, semiconductor saturable absorber mirrors, and ion-doped crystals have been widely used in commercial short-pulsed lasers. However, they suffer from limitations with respect to narrow wavelength ranges, slow recovery times, low flexibility in tuning the SA parameters, and high fabrication costs[114]. In contrast, 2D materials that feature broadband absorption and ultrafast recovery times are proving to be excellent SA materials[115]. Since the first demonstration of mode-locked fiber lasers based on SA in graphene in 2009[116], there has been an explosion in activity in using SA in 2D materials for short-pulsed lasers. The fast carrier relaxation times in 2D materials has also allowed the dynamic SA process to be used for ultrafast all-optical modulation, typically via pump-probe schemes[117,118].

Despite its relatively weak SA compared to PG, mainly due to its non-zero material bandgap, GO has been used for many pulsed fiber lasers because of its simple fabrication processes and the ability to tune its SA properties[104,119-121]. The SA in GO can arise from not only band-to-band transitions but also from defect states. For example, SA originating from the ground-state bleaching of the $sp^2$ orbitals has been observed[122,123], which had an energy gap of ~0.5 eV.

The principle of SA in GO is schematically illustrated in in **FIG. 6c,** showing fundamental similarities with other 2D materials such as PG and TMDCs. By leveraging both the SA and polarization selectivity of a GO-coated side-polished fiber in a fiber ring laser, stable mode-locked pulses (**FIG. 6d**, **left panel**) from a mode-locked fiber laser (MLFL) have been generated[103]. Switchable operation from mode-locking to Q-switching was demonstrated by changing the polarization state. A saturable absorber using rGO also resulted in highly stable mode-locked pulses from a MLFL[104] with a repetition rate of ~16.79 MHz and pulse duration of ~1.17 ps (**FIG. 6d**, **right panel**). Recently, SA in GO-coated silicon waveguides was investigated[106], showing increasing efficiency and decreasing power threshold with increasing film thicknesses.

### *Second-order optical nonlinearity*

The nonlinear response of a material excited by an external optical field (***E***($t$), scalar) can be expressed as (in scalar form for brevity, and in the dipole approximation)[97]

$$\boldsymbol{P}(t) = \varepsilon_0 \, (\chi^{(1)}\boldsymbol{E}(t) + \chi^{(2)}\boldsymbol{E}^2(t) + \chi^{(3)}\boldsymbol{E}^3(t) + \ldots) \tag{1}$$

where ***P***($t$) (scalar) is the light induced polarization, $\varepsilon_0$ is the vacuum permittivity, and $\chi^{(i)}$ are the $i^{\text{th}}$-order optical susceptibilities, which generally are tensors of rank ($i + 1$). In Eq. (1), $\chi^{(1)}$ describes the linear optical properties such as $n$ and $k$, while $\chi^{(2)}$ corresponds to 2$^{\text{nd}}$ order nonlinear optical processes including second-harmonic generation (SHG, **FIG. 6e**), sum /

difference frequency generation, the Pockels effect, and electro-optic rectification[99]. In the dipole approximation, the 2nd order nonlinearity is zero in bulk centrosymmetric materials such as silica and silicon, although it exists at the surface where the inversion symmetry is naturally broken, or at the interface between different materials[124-126]. Note that the susceptibilities in **Eq. (1)** are complex, with the real and imaginary parts contributing to different effects. For example, the real part of $\chi^{(1)}$ accounts for linear refractive index, and the imaginary part of $\chi^{(1)}$ is responsible for linear optical absorption.

The 2nd order optical nonlinearity has many applications in both the classical and quantum regimes, exemplified by optical autocorrelators[127], electro-optic modulators[128], optical coherence tomography[129], and quantum light sources[130]. Many 2D materials that lack inversion symmetry have displayed $\chi^{(2)}$ values much higher than bulk materials[131,132]. Unlike PG that is centrosymmetric, and hence has no $\chi^{(2)}$ unless induced by external excitations that break the inversion symmetry, GO can have a high 2nd order optical nonlinearity arising from its highly heterogeneous structure. Efficient SHG from GO has been observed[105], where a large hyperpolarizability of $(1.36 \pm 0.15) \times 10^{25}$ esu at 800 nm was obtained by measuring the molar mass distribution of 2D GO sheets dispersed in an aqueous suspension (**FIG. 6f**). Voltage-controlled SHG in a rGO/silicon heterointerface was also demonstrated when excited by a 1064-nm pulsed laser[133], offering new possibilities for integrated photonic devices with a tunable 2nd order optical nonlinearity.

## *Third-order optical nonlinearity*

In **Eq. (1)**, $\chi^{(3)}$ describes the 3rd order nonlinear optical processes that are present in all materials, in contrast to $\chi^{(2)}$ that only exists in non-centrosymmetric materials as mentioned. The 3rd order nonlinear response gives rise to a rich variety of processes, including four-wave mixing (FWM), self-phase modulation (SPM), cross-phase modulation (XPM), third-harmonic generation (THG), two-photon absorption (TPA), SA, stimulated Raman scattering, and many others. The response represented by **Eq. (1)** is virtual, or parametric – in other words, it is a fixed and fundamental property of the material band structure and operates through virtual excitations. However, processes that involve optical absorption, such as TPA or SA, can result in real photogenerated carriers which alter the quiescent material nonlinear response and render **Eq. (1)** incomplete. Real carrier effects are complex processes not only generated by, but can induce their own, TPA or SA. They tend to be much slower than the intrinsic virtual response since they depend on carrier diffusion, recombination, etc. The third-order parametric processes arising from $\chi^{(3)}$, in contrast, operate by virtual excitation of carriers without creating photogenerated carriers (**FIG. 6g**), and so are quasi-instantaneous with ultrafast response times on the order of femtoseconds[91]. This has motivated ultrafast all-optical signal generation and processing for telecommunications, spectroscopy, metrology, sensing, quantum optics, and many other areas[89,96].

Nonetheless, despite the success of nonlinear optical devices in optical communications and information processing[97,98,134], particularly those in integrated form, they are subject to continual research to improve their performance. For example, although silicon is a leading material for integrated optics, it has very strong TPA in the near-infrared regime, which limits its nonlinear optical performance in the telecom band. Other CMOS-compatible materials such as silicon nitride (SiN) and high-index doped silica glass (Hydex)[98,135] have a much weaker TPA, although this comes at the expense of a much lower Kerr nonlinearity than silicon.

In the past decade, the giant third-order nonlinear optical response of 2D materials has been widely recognized[99,113]. Although GO has a Kerr coefficient $n_2$ that is typically $10^{-14} - 10^{-13}$ $m^2\,W^{-1}$, which is lower than PG (typically $10^{-13} - 10^{-12}\,m^2\,W^{-1}$), it is nonetheless still 4 orders of magnitude larger than silicon. This is more than enough for practical devices such as coated optical fibers or integrated photonic chips. The large bandgap of GO also yields much lower linear light absorption and TPA in the near-infrared regime than PG, which is very attractive for nonlinear optical processes where high light intensities are typically used to enhance the nonlinear response. The nonlinear optical absorption arising from $Im(\chi^{(3)})$ processes such as SA and TPA is dependent on the bandgap of the prepared GO as well as the wavelength of the excitation light. Since there are different mechanisms and bandgaps for SA and TPA in GO, they can and often do coexist, leading to very complex and wavelength dependent nonlinear absorption.

Enhanced FWM in Hydex and SiN waveguides integrated with 2D GO films has recently been demonstrated[101,136], with an improvement in the conversion efficiency of ~6.9 dB for a uniformly coated Hydex waveguide and ~9.1 dB for a SiN waveguide with a patterned film. The deposition and patterning of the 2D GO films were realized by layer-by-layer self-assembly together with lithography and lift-off. These fabrication methods also yielded patterned 2D GO films integrated onto Hydex MRRs[53], where the resonant enhancement significantly increased the optical intensity in the cavities, increasing the conversion efficiency by ~10.3 dB (**FIG. 6h, left panel**). In addition to FWM, enhanced SPM in GO-coated silicon waveguides was also demonstrated. As compared with Hydex and SiN waveguides, silicon waveguides have smaller dimensions and consequently a much higher mode overlap with GO. The hybrid waveguides showed significantly improved spectral broadening of optical pulses (**FIG. 6h, right panel**) compared to uncoated silicon waveguides, achieving a high broadening factor of 4.34 for a 3-mm-long waveguide integrated with a patterned GO film.

## Electronic properties

The electrical conductivity of graphene based materials can be efficiently modulated by varying the Fermi energy level and material bandgap. It has been observed that GO can transit from an insulator to semiconductor and then to a semi-metal during the reduction process, as the bandgap decreases (**FIG. 7a**)[137]. Thus, the range in tuning the material transition from GO to trGO is much larger than what can be achieved by just doping PG[138]. On the other hand, trGO is very close to a gapless semimetal – with a Fermi energy level at the Dirac point of the bandstructure that can shift down or up through chemical doping (**FIG. 7a**), thus yielding p-type or n-type doped rGO. The doping process produces a bandgap between the valence and conduction bands that converts the semi-metal graphene to a semiconductor. Furthermore, doping modifies the molecular structure, which affects the electrical, mechanical, optical, and thermal properties of doped rGO.

Chemical doping can be categorized into: 1) substitutional doping, or 2) surface transfer doping. Here, we mainly focus on substitutional doping by exchanging carbon atoms with heteroatoms to tune the Fermi energy level of rGO, where covalent bonds link the doped atoms to the carbon network. This approach has been used to realize metalloids and non-metals incoporating B, N, P, S, F, Cl, and Si, which form either p-type or n-type doping, depending on the number of valence electrons of the dopant. For example, since B has fewer valence electrons, it produces holes in valence band, thus downshifting the Fermi energy level to form p-type doping (**FIG. 7b**). In comparison, N and P provide extra electrons in the conduction band, thus upshifting the Fermi energy level to from n-type doping. As a result, the conductivity of rGO can be tuned by adjusting the type and concentration of the doping.

The ability to tune its electrical properties has enabled many electronic devices based on rGO. Combined with the ability to pattern the material conductivity via laser photo reduction, this has been used for microheater electrodes (**FIG. 7c**)[40]. By reducing the thickness of rGO to decrease its light absorption, patterned rGO can also serve as transparent conductive electrodes for light emitting diodes (LEDs) (**FIG. 7d**)[139] and solar cells[140]. For LEDs, since only the laser exposed areas are conductive, it enables patterned lighting. Further, it is also possible to achieve incandescent light emitters based on rGO "papers" composed of monolayer rGO and single walled CNTs (**FIG. 7e**). This system can withstand high temperatures up to 3000 K[141] – substantially higher than pure CNTs or rGO, and most importantly, with a lighting efficiency comparable to tungsten wires. Due to the ultrathin freestanding design, these rGO "papers" are bendable and lightweight, useful for wearable electronics and fast heating.

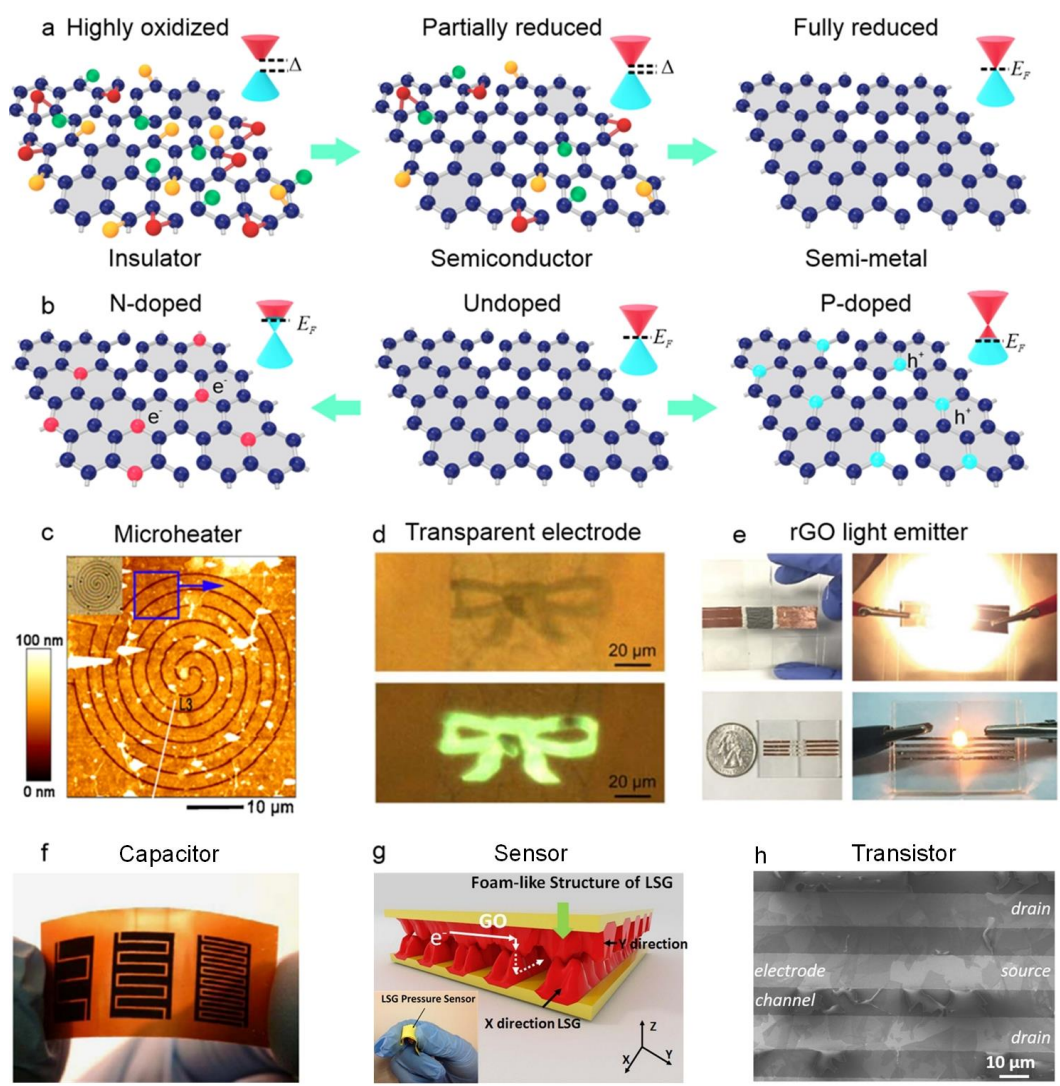

**Fig. 7. Electronic properties and applications. a**|Tuning the bandgap of GO by reduction. **b**| Tuning the Fermi energy level of rGO by different types of doping. **c**| AFM image of a rGO microheater. **d**|Optical microscopy image of rGO used as transparent electrode on LEDs. **e**| Photos of a rGO light emitter. **f**| A photo of a GO/rGO supercapacitor. **g**| 3D conceptual drawing and photo of a rGO sensor. **h**| SEM image of a rGO transistor. Panel **c** adapted with permission from REF.[40], Elsevier Publishing. Panel **d** adapted with permission from REF.[139], ACS Publications. Panel **e** adapted with permission from REF.[141], Wiley-VCH. Panel **f** adapted with permission from REF.[142], Springer Nature Limited. Panel **g** adapted with permission from REF.[143], Springer Nature Limited. Panel **h** adapted with permission from REF.[144], Springer Nature Limited.

Porous rGO has a high conductivity and large surface area that are both ideal for electrochemical energy storage (**FIG. 7f**)[142]. For laser-patterned GO films, the unexposed or unreduced areas remain insulating, which can be used as an in-plane electronic insulator for interdigital supercapacitors that have achieved high power and energy density[10]. The porous structure also exhibits a significant change in conductivity as a result of applied external pressure, which has been used for realizing pressure sensors (**FIG. 7g**) with high sensitivities over a wide range of pressure[143]. In addtion, the high stability and mechanical strength of GO as well as its low electrical conductivity have been exploited for anti-self-discharge seperation in lithium-sulfur batteries[145].

Recent progress in microwave reduction methods have shown that the electrical properties of rGO can be made extremely close to that of PG, and so field effect transistors (FETs) based on rGO have been demonstrated[11] and used for detecting human papillomavirus (HPV) (**FIG. 7h**)[144], proving the viability of these devices for clinical applications.

## Optoelectronic properties

External electrical fields can not only alter the optical response of GO and rGO, but the reverse is true – light excitation can modify their electrical response, and this interplay forms the basis for optoelectronic devices. In addition to distinct optical and electronic properties, GO and rGO also exhibit remarkable optoelectronic properties. In this section, we summarize them and review their applications.

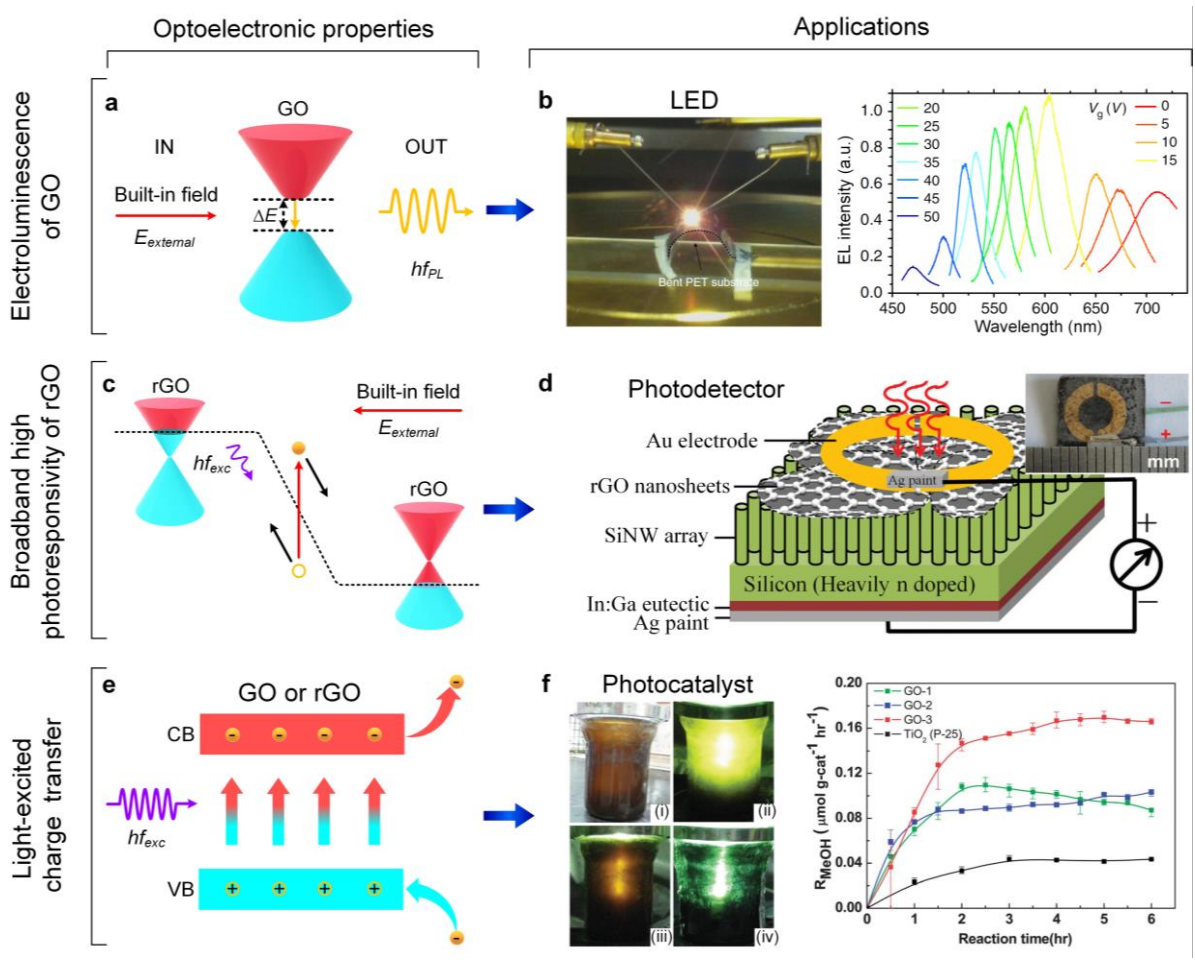

Fig. 8. **Optoelectronic properties and applications. a|** Schematic band diagram for electroluminescence process in GO. **b|** Light emission from a GO-/rGO-based light-emitting diode (LED) (left) and luminescence spectra for varied gate bias from 0 to 50 V (right). **c|** Schematic showing photocurrent generation in rGO induced by an external electrical field. **d|** Schematic illustration of a photodetector (PD) based on rGO-silicon nanowire array (SiNW) heterojunctions. Inset shows a photograph of the fabricated device. **e|** Schematic illustration of light-excited charge transfer in GO or rGO. CB: conduction band. VB: valence band. **f|** Left: color variation of GO suspension during photocatalytic reaction under mercury-lamp irradiation: i) before irradiation, ii) start of irradiation, iii) irradiation for 30 min, iv) irradiation for 2 h; Right: Photocatalytic methanol formation rate (RMeOH) on different GO samples (GO-1, GO-2, GO-3) and $TiO_2$ under visible light irradiation. Panel **b** adapted with permission from REF.[146], Springer Nature Limited. Panel **d** adapted with permission from REF.[147], Wiley-VCH. Panel **f** left adapted with permission from REF.[148], Wiley-VCH; Panel **f** right adapted with permission from REF.[149], RSC Publishing.

## *Electroluminescence*

Electroluminescence, where electrons are excited by external or build-in electrical fields and subsequently radiate photons when they return to the low energy state, is similar to fluorescence except with a different excitation mechanism. A typical example of electroluminescence is LEDs, which are key components in the display industry. The light-emitting mechanism of LEDs is different to that of incandescent light sources such as tungsten filaments and the rGO light emitter mentioned in **FIG. 7e**, which can result in significant benefits in energy consumption, lifetime, device footprint, and switching speed.

The large and tunable bandgap of GO, together with its tunable electrical conductivity and high thermal conductivity, has been used to realize high-performance LEDs. A GO / rGO-based field-effect LED[146] is shown in **FIG. 8b**. The light emission was based on semi-reduced GO formed at the rGO / GO interface created via laser scribing. The rGO nanoclusters in the semi-reduced GO interfacial layer have a range of sizes, resulting in different discrete energy levels that allow selective excitation of luminescence by electrical gating or changing the doping level. Tunable electroluminescence from blue to red was achieved, with a brightness of 6,000 cd m$^{-2}$ and an efficiency of ~1 %. Further, the high thermal conductivity of rGO has been exploited for heat removal in integrated gallium nitride (GaN) LEDs[150], which significantly improved the heat dissipation performance, resulting in a much lower junction temperature and thermal resistance compared to devices without GO.

*Photovoltaic effect*

The photovoltaic effect is the generation of voltage or current in an irradiated material. The principle of photovoltaic photocurrent generation in rGO is schematically illustrated in in **FIG. 8c**, where the photo-generated electron–hole pairs are separated by an external electrical field.

An important application of the photovoltaic effect is photodetectors (PDs), particularly in the context of 2D materials[151]. PDs play a key role in optical communications, sensing, imaging, astronomical detection, and environmental monitoring[151,152]. Although conventional PDs based on bulk materials (e.g., Ge, Si, and InGaAs) are mature, their response speed, detection efficiency, and operation wavelength range can always be improved to meet ever-increasing performance demands. Graphene-based PDs have attracted significant interest over the past decade because of their wavelength-independent absorption arising from the zero bandgap as well as their ultrafast response enabled by high carrier mobility[153-157]. This has resulted in superior performance compared to bulk-material-based PDs. In contrast to unreduced GO where the light absorption and electrical conductivity are both too low for PDs, trGO has material properties that are close to graphene, which have enabled many rGO-based PDs with performance comparable to graphene-based PDs[147,158-170]. In addition, rGO has unique advantages for mass producibility, compatibility with diverse substrates, and flexibility in material engineering, making it attractive for manufacturable PDs.

A PD implemented by rGO-silicon nanowire array formed by self-assembly[147] is shown in **FIG. 8d**. The broadband response of rGO resulted in a PD operating from the visible to the terahertz (THz) regime, reaching a maximum photoresponsivity of 9 mA/W in the mid-infrared regime. With a high sensitivity to detect mid-infrared radiation from the human body, this PD is attractive for human infrared sensors. Another rGO PD has achieved an ultrahigh photoresponsivity of $3.1 \times 10^4$ A/W, together with a remarkable external quantum efficiency (EQE) of $1.04 \times 10^7$ %[166]. The photoresponsivity was about four orders of magnitude higher than comparable graphene devices, with the EQE outperforming commercial silicon PDs. This high performance resulted from the specially designed 3D nanoporous structure of rGO, which showed > 40 times higher light absorption than graphene, together with an electron mobility that was over a hundred times higher than rGO films formed from nanoflakes.

The photovoltaic effect has also been utilized in solar cells that convert solar energy into electrical power[171], in contrast to solar absorbers mentioned previously that use solar thermal energy to drive heat engines. Polymer composites incorporating functionalized GO sheets have been employed as a photoactive layer in organic photovoltaic cells[172], achieving a maximum energy conversion efficiency of 1.4 %. Photoactive layers formed by polymers incorporating rGO quantum dots were also investigated, showing significantly improved photovoltaic performance compared to photoactive layers formed by rGO sheets[173].

## *Photoconductivity*

Photoconductivity, also termed the photogating effect[151], is based on light-induced charge transfer (**FIG. 8e**) that changes the carrier density and consequently the material conductivity. An important application of the photoconductive effect is photocatalysis, which has a long history and has experienced significant development in the past 50 years. Photocatalysts drive chemical reactions by absorbing light to induce charge transfer, and so GO, with its broadband absorption and excellent semiconductor properties, is attractive as a nontoxic photocatalyst with high efficiency.

The use of GO as a semiconductor photocatalyst for producing hydrogen via water dissociation has been achieved[148], where moderately oxidated GO steadily catalyzed the generation of hydrogen from an aqueous methanol and pure water solution under ultraviolet or visible irradiation (**FIG. 8f, left panel**). Photocatalytic conversion of carbon dioxide to methanol based on GO has also been realized[149], with a high conversion rate of 6 times that of titanium dioxide ($TiO_2$), a conventional photocatalyst, under visible irradiation (**FIG. 8f, right panel**).

## Other applications

Either in its pristine form or a reduced form, GO has underpinned many other technologies related to photonics, electronics, and optoelectronics. They generally leverage GO's chemical or physical properties such as its high hydrophilic nature, strong capillary action, high surface-to-volume ratios, and the electronic and optical properties mentioned above. Some of these technologies have already become widely successful on a large scale and are summarized in **Table 1.** We discuss these next.

Photothermal therapy is commonly used in cancer treatment with minimal invasiveness, high specificity, and low toxicity. It relies on the activation of photothermal agents taken up by cells via light absorption, thus generating heat that kills cancer cells via thermal ablation. The photothermal agents usually absorb strongly in the near-infrared regime whereas normal cells in healthy tissues do not. With strong near-infrared absorption and high photothermal efficiency, GO and rGO in the form of nanoparticles have been demonstrated to be high performance photothermal agents[174-176]. As abundant materials that can be synthesized from graphite, they are also inexpensive compared to other photothermal agents such as gold-based nanoparticles[177]. Polyethylene glycol (PEG) coated GO was first used for in-vivo photothermal therapy[174], achieving ultra-efficient tumor ablation in mice without obvious side effects. Subsequently, rGO nanosheets with noncovalent PEGylation, which had more than 6 times higher absorption in the near-infrared regime than nonreduced, covalently PEGylated GO, were employed for selective photoablation of cancer cells[175]. Photothermal treatment of Alzheimer's disease based on functionalized GO has also been achieved[176], where GO modified by thioflavin-S (ThS) effectively dissociated amyloid deposits in mice cerebrospinal fluid after near-infrared irradiation.

**Table 1 | Summary of other typical applications of GO and rGO.**

| Materials | Applications | Materials properties | Main feature | Ref. |
|---|---|---|---|---|
| GO | Photothermal therapy | 1. Strong optical absorption to generate local heat<br>2. High dissolubility in biomedium | Remarkably reduced side effects and improved selectivity compared to chemotherapies and radiotherapies | 176 |
| rGO | Protective coatings | 1. Optically transparent<br>2. Extremely low permeation of gases and liquids for multilayer rGO films<br>3. High stability and little toxicity<br>4. Solution-based fabrication for large-scale manufacture | 1. No detectable permeation of hydrogen and moisture for a 30-nm-thick films<br>2. No detectable permeation of all tested gases and liquids for the films thicker than 100 nm | 178 |
| GO | Molecular and Ionic sieving | 1. Well-defined nanometer pores<br>2. Low frictional water flow<br>3. High surface-to-volume ratio<br>4. High pressure inside the capillaries | 1. Blocking solutes with radii > 4.5 Å<br>2. High permeation rate thousands of times faster than simple diffusion | 179 |
| GO | Antibacterial agent and nanotoxicology | 1. Strong dispersion interactions with lipid molecules<br>2. Destructive extraction of phospholipids from the cell membranes to reduce bacteria viability. | Green antibacterial material with little bacterial resistance | 180 |
| GO | Paper production | Unique interlocking-tile arrangement of GO nanosheets in a near-parallel fashion enabled by directed-flow assembly | 1. Outperforming traditional carbon- and clay-based papers in stiffness and strength<br>2. With chemical tunability for diverse applications | 16 |

Protective coatings to guard against corrosion, mechanical wear, fouling, and microorganisms are integral to many industries such as food, medicine, electronics, chemicals, and military[181]. In contrast to conventional metal coatings or organic paint that often have hazardous constituents with negative environmental impact, GO and rGO can serve as low-cost and eco-friendly protective coatings that also have excellent gaseous impermeability, anti-bacterial property, adsorption capacity, visible light transparency, mechanical strength, and thermal stability. High performance impermeable barrier films based on chemically reduced GO films have been reported[178]. For the 30-nm-thick films that are optically transparent, no hydrogen or moisture permeation was detected, while for > 100 nm thick films, all tested gases, liquids, and aggressive chemicals were blocked with no detectable permeation.

Molecular and ionic sieving based on thin membranes that allows fast separation of solutes in aqueous media has a key role in water purification and desalination as well as substance detection and filtration. Ultrasmall GO membrane pores on nano or sub-nano scales display a high capillary pressure. This, along with the large surface-to-volume ratios, is extremely effective as molecular and ionic sieves[179,182-184]. GO membranes with a thickness of 1.8 nm achieved extremely high hydrogen separation selectivity of 3400 and 900 for $H_2/CO_2$ and $H_2/N_2$ mixtures, respectively[182]. Molecular sieves based on micrometer-thick GO laminates were also demonstrated, which blocked all solutes with radii >4.5 angstroms in water, at permeate rates thousands of times higher than normal diffusion[179]. Subsequently, zeolitic imidazolate framework-8 (ZIF-8)-nanocrystal-hybridized GO membranes were demonstrated to have an ultrahigh water permeability of 601 $m^{-2}$ $h^{-1}$ $bar^{-1}$ – a 30-fold increase over comparable pure GO membranes[18]. The strong molecular sieving capability of GO membranes was also used for highly sensitive substance detection based on the changes in GO's optical or electrical properties after absorption of target gases or organic compounds[183,185]. In addition, reversible fusion and fission in GO fiber shells induced by solvent evaporation and infiltration were observed recently[186], making GO an attractive material for constructing dynamically transformable systems.

Biomedicines for drug delivery, disease diagnosis, and controlled drug release are an important application of nanomaterials. In the past decade, the excellent antibacterial capability of GO and GO composites has been widely recognized. It has been demonstrated that GO exhibits better antibacterial performance than rGO and graphite[187]. More importantly, since its antibacterial action is based on physical damage, GO based antibacterial agents have low toxicity and little bacterial resistance. It has been observed that GO nanosheets can interfere with bacteria and disrupt their cell membranes[188]. It is also found that GO nanosheets can also induce destructive extraction of phospholipids from *Escherichia coli* and consequently greatly reduce their viability[180].

Finally, GO has also been widely exploited for paper production[16,189]. GO papers made from GO nanosheets in an interlocking-tile arrangement have a significantly improved mechanical stiffness and strength as compared with traditional papers[16]. The mechanical strength of GO paper materials can be improved even further by modifying GO to incorporate divalent ions such as $Mg^{2+}$ and $Ca^{2+}$[190]. The combination of simultaneously achieving a free-standing paper with antibacterial capability has resulted in high-performance GO antibacterial papers[191]. Insulating GO papers can also be converted to electrically conductive rGO papers via thermal annealing or chemical reduction[192,193].

## Future prospects

We have reviewed recent advances in GO based photonics, electronics, and optoelectronics, clearly showing that this promising interdisciplinary field has progressed tremendously in the past decade. Despite this, though, there is nonetheless still significant room for new innovations and improvement. In this section, we conclude this Review by discussing these open challenges as well as future prospects for establishing certification standards and synthesis protocols, improving material quality, theoretical modelling, and opening up new applications. These issues are intimately connected, and their synergy will underpin future technogies based on GO photonics, electronics, and optoelectronics.

## *Certification standards and synthesis protocols*

As shown in **FIG. 9**, increasing the degree of reduction increases the electrical conductivity of rGO for a diverse range of electronic applications that require the conductivity to be varied by orders of magnitude. For example, transistors[144] and transparent electrodes[57] require the highest conductivities (typically ~$10^6$ S/m), whereas most sensors[143] and microheaters[40] only need a much lower conductivity. In comparison, when using GO as separators of structural capacitors[194,195] an extremely low conductivity is required to provide high electrical isolation. This also applies to interdigital microsupercapacitors where unreduced GO acts as in-plane seperators[142]. On the other hand, the optical properties of rGO, such as the refractive index *n* and extinction efficient *k*, both increase with the degree of reduction. Similar to electronics, photonic applications of GO require a range of properties. For example, solar absorbers[51] require high light absorption, and thus trGO is more suitable. The same is true for PDs[147,167] that require both high light absorption and conductivity to generate strong photocurrents. In addtion, a high absorption and/or refractive index contrast between the reduced and unreduced regions are beneficial for flat lenses[32,60-62] and holograms[74,196] to widely vary the amplitude and phase. Fluroscence[148] and photocatalyst[51] applications mainly utilize the semiconductor properties of semi-reduced GO. Finally, in order to minimize the loss of waveguide devices or spacial-light polarizers[12,26], highly oxidized GO films are critical since they are pure dielectrics with minimal absorption. In summary, there is not one set of criteria suitable for all applications, but rather the entire family of graphene based materials is highly useful due to their high flexibility and tunable properties. GO and rGO signfiicantly increase the variety of graphene based technologies and devices that can be realized, well beyond that of simple PG.

The recent growth in the field of GO based devices has significantly advanced industrial scale applications of graphene family materials. To this end, establishing large-scale cost-effective synthesis methods and widely recognized certification standards are critical. The molecular structure of GO films can be quantified by a few key parameters that include the C-

O ratio, the $I_D/I_G$ and $I_{2D}/I_G$ ratios, flake sizes, and layer numbers. The C-O ratio, which can be measured via XPS, is a key parameter to quantify the reduction degree of GO[11,67] – the lower the C-O ratio, the lower degree of reduction. Second, the $I_D/I_G$ and $I_{2D}/I_G$ ratios, which can be obtained from Raman spectroscopy[11], are widely used to characterize the properties of graphene family materials. The $I_D/I_G$ ratio provides a good indication of the defect density of GO/rGO materials – the lower the ratio, the lower the defect density. The $I_{2D}/I_G$ ratio reflects the formation of graphene domains – the higher the ratio, the closer to PG. Lastly, the flake sizes in the GO films can be measured via SEM and the layer numbers determined by measuring the film thicknesses via AFM[17].

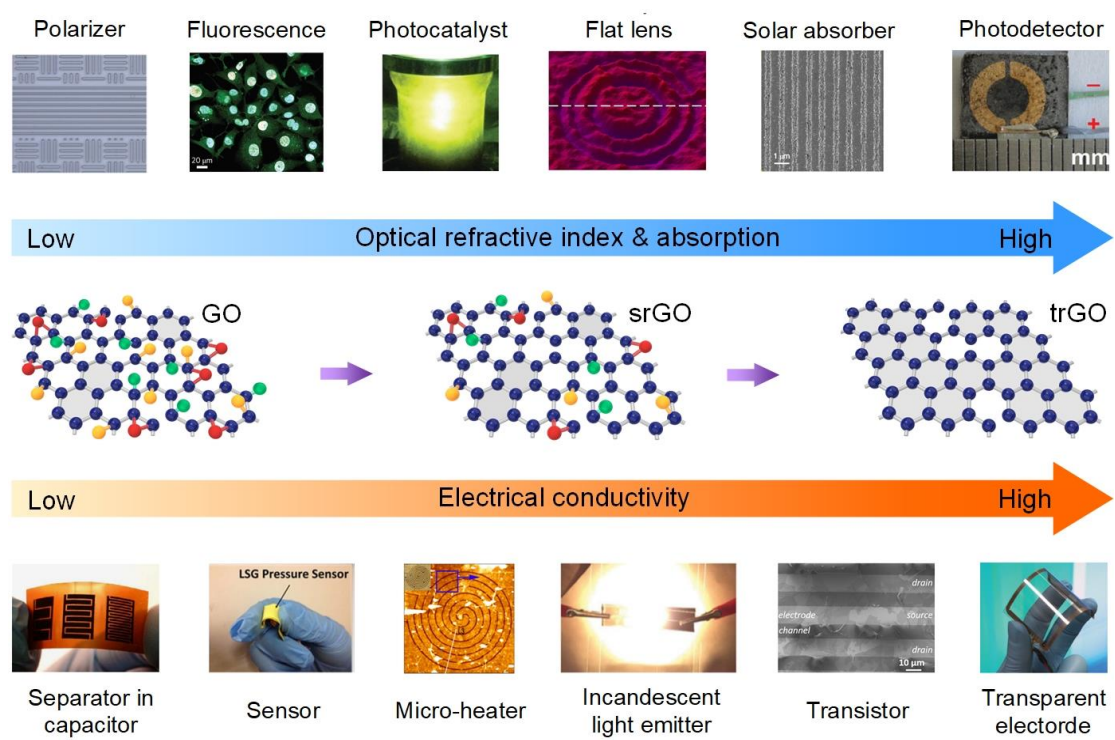

Fig. 9. **Diverse applications of GO with different degrees of reduction.** srGO: semi-reduced GO. trGO: totally reduced GO. Insets in the top row from left to right are adapted with permission from REF.[12] (Wiley-VCH), REF.[102] (Springer Nature Limited), REF.[148] (Wiley-VCH), REF. [51] (Springer Nature Limited), REF.[51] (Springer Nature Limited), and REF.[147] (Wiley-VCH), respectively. Insets in the bottom row from left to right are adapted with permission from REF.[142] (Springer Nature Limited), REF.[143] (Springer Nature Limited), REF.[40] (Elsevier Publishing), REF.[141] (Wiley-VCH), REF.[144] (Springer Nature Limited), and REF. [57] (Springer Nature Limited), respectively.

For applications requiring both low electrical conductivity and optical absorption, it is preferable to use GO films with a low C-O ratio, in the range of 1:4 to 1:1[32,36,61], which also results in high $I_D/I_G$ and low $I_{2D}/I_G$ ratios. On the other hand, for applications requiring a high electrical conductivity[57,144] or optical absorption[51,197], the requirements are the opposite. The measured film electrical conductivity and optical absorption can also be used to quantify the degree of reduction of the GO material. In addition, in order to fabricate smooth GO/rGO films, monolayer GO flakes with sizes on the order of 100's nm are needed[17]. For more specialized applications, such as conformal coating over nanostructured surfaces including slot waveguides[198] or any nanostructures with a small gap size (< 100 nm), the GO flake size should be reduced even further[70].

Since the material properties of GO and rGO are affected by the fabrication methods, the quality and consistency of synthesized materials in practical settings vary widely, which has limited its large-scale industrial applications. In order to overcome this challenge, it is necessary to develop synthesis protocols. Liquid phase exfoliation is attractive for mass producing GO flakes from natural graphite, which can produce GO in quantities of tons[199]. The exfoliated GO flakes can be further used for either solution based applications or for film coatings. The reaction during the exfoliation process needs to be properly controlled in order to obtain the desired degree of reduction, which can be quantified by the C-O ratio as mentioned before. So far, the most effective method for producing highly oxidized GO flakes is the modified Hummer's method[24], which is able to produce monolayer flakes with a high oxidation extent of up to 69 % .

For thin films, the synthesis protocols should include two addtional steps – film synthesis and GO reduction. To synthesize GO films in industrial scale, it is necessary to achieve fast, large-scale, and roll-to-roll production. For energy storage and most electrical applications that do not require accurate thickness control, the Dr. Blade method allows roll-to-roll production of GO films with thicknesses of a few microns[27]. On the other hand, most photonics based applications require very accurate thickness control down to a few nanometers (~1/100 of the optical wavelength). In this case, the appropriate approach is the solution based layer-by-layer self-assembly method[12,51,53].

For the reduction of GO films, methods based on electromagnetic wave exposure (using light or microwaves) usually produce higher quality films than thermal or chemical reduction methods, and are also fast and efficient. For large-scale homogeneous reduction, microwave methods produce the material with the highest quality that approaches PG, but suffer from limitations in terms of film patterning and control of the degree of reduction. In contrast, photoreduction methods, including flashlight[67] and laser reduction[37], show excellent capabilities in these aspects, although they have thus far not been able to achieve the same level of material quality as that of microwave reduction. Therefore, in principle the combination of both microwave and photoreduction methods may yield the optimal approach. Indeed, microwave reduction requires a pre-reduction[11] to increase the conductivity of the GO films, and so in principle it would be feasible to use either laser or flashlight exposure to complete the pre-reduction process instead of the typically used hot plate method. This would allow rGO devices to be fabricated and patterned with a very high film quality approaching that of PG. Alternatively, the process can be reversed, with laser ablation[51] to post-pattern large areas of microwave reduced GO films, which combines the advantages of microwave reduction with laser ablation patterning. Conventionally, laser patterning is carried out in a direct writing, or point-by-point manner, by scanning a single focal spot, thus signficantly limiting the overall processing speed. The development of multifocal parallel laser printing[200] can significantly increase the speed by more than two orders of magnitude, making it attractive for industrial scale applications. Finally, these hybrid approaches, based on a combination of both optical and microwave reduction methods, are energy efficient, fast (on a timescale of a few seconds) and green (not involving or producing any toxic chemicals).

In summary, we believe that the most attractive methods for GO based device fabrication will involve synthesis protocols that combine the modified Hummer's method with roll-to-roll film synthesis, including Dr. Blade coating and layer-by-layer self-assembly, together with film reduction and patterning methods that combine the advantages of optical and microwave techniques.

## *Material quality*

**Table 2** compares the state-of-the-art performance of GO, rGO, and graphene devices for a range of applications including electronic, photonic, and optoelectronic devices that incorporate solid films. For solution-based applications such as fluorescence or photocatalysis, GO and properly reduced GO show much better performance than graphene due to their higher solubility and better ability to be processed in water and polar solvents. In **Table 2**, we see that the carrier mobility of rGO has improved considerably in the last 10 years[201], with current results being only about an order of magnitude lower than PG. Further improvement would make rGO more suitable for FETs and transparent electrodes. To date, for many uses such as solar light absorption, polarization selection, mode-locked lasers, thermal harvesting, and photo detection, GO and rGO devices have already shown comparable or even better performance than graphene devices. With its highly heterogeneous structure, GO outperforms centrosymmetric graphene in areas that are based on second-order nonlinear optical processes. For applications based on third-order nonlinear optical processes, they each have advantages and disadvantages. Although GO has a smaller optical nonlinearity than graphene, its significantly lower linear loss is very advantageous for nonlinear optical processes, and its lower nonlinear loss (e.g., two-photon absorption) is also beneficial at the high light intensities typically employed in these devices. The adjustable reduction of GO provides a way to properly balance the trade-off between optical nonlinearity and linear / nonlinear loss in order to optimize the device performance.

**Table 2 | Comparison of the state-of-the-art performance of GO, rGO, and graphene devices for electronic, photonic, and optoelectronic applications.**

| Application | Key performance parameter | GO | rGO [a] | graphene | Refs. |
|---|---|---|---|---|---|
| Electrodes or FETs | Carrier mobility ($cm^2 V^{-1} s^{-1}$) | ~ 0 | ~ $10^3$ | ~ $10^4 - 10^5$ | 5,11,22,202 |
| Solar absorbers | $k$ [b] | <0.01 | 1.95 | 2.0 | 12,17,51 |
| Broadband polarization selective devices | Extinction ratio (dB) Operation band (nm) | 53.8 632–1600 | – [c] | 27 488–1650 | 12,75 |
| 2nd order nonlinear optical processes | $\chi^{(2)}$ (esu) | $1.36 \times 10^{-25}$ | – [c] | ~ 0 | 99,105 |
| 3rd order nonlinear optical processes | $n_2$ ($m^2 W^{-1}$) [b] | $1.5 \times 10^{-14}$ | $-5.4 \times 10^{-14}$ | $-8.0 \times 10^{-14}$ | 101,136,203 |
| Mode-locked lasers and all-optical modulation | Nonlinear absorption coefficient ($m W^{-1}$) [b] | $-6.2 \times 10^{-8}$ | $4.3 \times 10^{-8}$ | $9.0 \times 10^{-8}$ [d] | 106,203,204 |
| Thermal harvesting and heat removal | Thermal conductivity ($W m^{-1} K^{-1}$) | 8.8 | 2000 – 4000 | ~ 5000 | 28,205,206 |
| Broadband PDs | Photoresponsivity ($A W^{-1}$) [e] Operation band (nm) | N. A. [f] | ~ 1.5 365 – 1200 | ~ 1.0 500–3200 | 156,167 |

[a] Here we show the state-of-the-art values for highly or totally reduced GO.
[b] Here we compare typical measured values at 1550 nm.
[c] There are no reported works on this so far.
[d] Here we show the measured value for undoped graphene. The nonlinear absorption of graphene can be changed by doping or introducing defects.
[e] Here we show and compare the performance of chip-scale integrated PDs incorporating rGO and graphene.
[f] N.A.: not applicable. GO is a dielectric that cannot be directly used for photo detection.

Given the close relationship between GO and graphene, the efficient production of high-quality graphene from mass-produced GO is fundamental to many applications. This has been a driving force since the discovery of graphene, but it is only in the past decade that exciting progress has been made, with the gap between the properties of rGO and PG being greatly reduced by the methods based on wet chemistry[22] and microwave reduction[11]. Advancing these methods further to achieve high-efficiency production at scale, together with precise patterning of the reduction area and fine control of the reduction degree, will be key subjects of future work. The first is important for the large-scale manufacture of GO devices for industry, while the latter two are critical for optimizing the material properties and device performance.

Reducing the linear optical absorption of GO is critical for many applications, particularly those based on nonlinear optical processes such as the Kerr effect, FWM, and THG. Theoretically, one might expect that there is no linear light absorption in the near infrared regime for pure GO since it has a bandgap > 2 eV. However, there are several factors contributing to the non-zero loss of practical GO films. These mainly include 1) scattering loss due to film surface roughness and imperfect contact between adjacent layers, and 2) linear light absorption induced by localized defects as well as foreign impurities. The important point, however, is that, since the linear loss from these sources largely stems from the material growth and fabrication processes, it is not a fundamental property. Hence, there is a significant potential to reduce the loss even further than the current state-of-the-art GO hybrid waveguides, that already have propagation losses that are two orders of magnitude lower than comparable graphene hybrid waveguides[12,53,106,136].

## *Theoretical modelling*

The exact chemical structure and fluorescence mechanisms of GO have been debated ever since the early work in this field[5] and this is still the case, although progress has been made, including new theoretical models that have been proposed to explain experimental observations[207-213]. The challenges mainly come from the non-stoichiometric heterogeneous structure of GO as well as its dependence on preparation processes. The heterogeneous structure of GO also enables the coexistence of different linear and nonlinear processes. Balancing these processes to optimize the performance for different applications requires a more in-depth understanding as well as more precise engineering of GO's material properties, which will be greatly assisted with better nano-characterization techniques.

Despite the advances in devices based on the nonlinear optical properties of GO and rGO, they are mainly semi-empirical and lack in-depth theoretical modelling. Many physical insights are yet to be explored, such as the source of anisotropy in either $n_2$ or the nonlinear absorption, the interplay between the different nonlinear absorption mechanisms (e.g., TPA and SA), the dispersion of the optical nonlinearity, and the interaction between $\chi^{(2)}$ and $\chi^{(3)}$ processes. Given its outstanding capability for achieving highly precise on-chip integration with mature fabrication technology, investigating GO in this form offers a promising avenue for investigating its material properties. In turn, this will enable integrated devices with new functionalities and improved performance.

In order to further control and optimize the material properties of rGO , it is necessary to understand and theoretically model the reduction process, such as how electromagnetic (either microwave or optical) waves interact with GO to reduce it. Empirically, molecular dynamic simulations can visualize the structure evolution, while first principles simulations can shed light on how the electromagentic energy is absorbed by specific bonds in order to dissociate them and remove the OFGs. The electromagnetic power, polarization, wavelength, optical pulse duration, or whether it is pulsed or continuous wave, are important parameters that govern this process.

Molecular dynamic simulations can simulate the structural evolution during thermal reduction by applying a static heat field[45], which predicts the final C-O ratio and molecular structure after reduction. To date, there have not been any molecular dynamic simulations of the reduction of GO films via microwave or photoreduction. The key challenges are how to apply a fast alternating electromagnetic field in the simulation and to accurately model how different covalent bonds respond to this field, given that different bonds have unique binding energies and so will respond differently. First principles simulations[214] have shown that extremely short femtosecond laser pulses (< 5 fs) can effectively remove OFGs without heating up the graphene network and thus generating defects. However, these ultrashort pulses are challenging to experimentally achieve and so it is of more interest to simulate the reduction process with standard pulse widths (e.g., 100 fs). Some theoretical and experimental work has demonstrated that the shape of femtosecond pulses can enable laser controlled chemical reactions to manipulate the end products[215]. Although laser controlled chemical reactions so far have only been used in gas reactions, the principle of controlling the reaction paths can be applied to the reduction of GO. This will open up new possibilities for the selective removal of particular OFGs, which can then be used to further tailor the properties of rGO films for new applications. So far this has only been acheived with conventional chemical reactions using toxic chemicals[216]. In order to realize this ambitious goal, an in-depth understanding of GO reduction via electromagnetic waves will be critical.

## *New applications*

Improving the performance of current devices will naturally open up new applications. For example, by increasing the level of SPM in GO hybrid waveguides, broadband super-continuum generation may be achievable. This could be realized by tailoring the waveguide dispersion for phase matching over a broader wavelength range. Another example is achieving broadband ultrafast all-optical modulation by using the SA of GO. Owing to the broadband SA response of GO, the pump and probe wavelengths can differ by hundreds of nanometers, which is extremely challenging for lithium niobate modulators or integrated silicon modulators that have typical operation wavelength ranges of < 100 nm[128,217]. For quantum optical applications, entangled photons generated from enhanced 2$^{nd}$ or 3$^{rd}$ order nonlinear optical processes such as SHG, THG, and FWM in GO based devices can be exploited to facilitate quantum optical information processing[95,218].

GO offers new possibilities for the combination of optical micro-combs and 2D materials to achieve new functionalities. Optical micro-combs are a key technology that has enabled many breakthroughs[90,219-221]. However, the challenges to achieve high parametric gain and self-starting and robust mode-locking significantly limit their performance for practical applications. Since the first demonstration of gate tunable optical micro-combs incorporating graphene[222], the micro-comb research community has recognized the strong potential of 2D materials to address these challenges. With its ultrahigh $n_2$ as well as the relatively low linear and nonlinear loss, GO is promising for enhancing the parametric gain and improving the comb conversion efficiency (**FIG. 10a**). The SA in GO could also be utilized for simple passive mode locking. Several key issues to address in future work include lowering the linear loss of GO films, accurate dispersion engineering, and achieving the proper balance in the trade-off between increasing the Kerr nonlinearity while limiting any increase in the linear loss.

Since GO has an ultrahigh optical nonlinearity, it can also be used for implementing high performance nonlinear optical fibers, which have applications in fiber pulsed lasers, sensing, and all optical signal processing[223]. The state-of-the-art in this area mainly implements GO films by coating optical fibers on their facets or laterally on planarized surfaces[103,104]. However, the nonlinear performance of these devices is limited by the weak light-GO interaction arising from either a short interaction length or weak mode overlap. Recently, the direct growth of 2D MoS$_2$ films onto the internal walls of silica optical fibers has been realized using capillary pre-deposition followed by chemical vapor deposition (CVD)[224]. This will inspire the future fabrication of nonlinear optical fibers with embedded GO and rGO (**FIG. 10b**), which would significantly enhance the light-GO interaction and hence the nonlinear performance. Given the excellent water solubility and self-assembled capability of GO nanoflakes, similar capillary filling methods can be used for coating GO films on the internal fiber walls. In this case the fabrication would be even simpler since it would not need subsequent CVD processes, and the coated GO films could be reduced further to form rGO.

In its 2D form, heterogeneous GO can form Van der Waals heterostructures that feature new properties[225,226]. By stacking 2D GO films with other 2D materials (**FIG. 10c**), charge redistribution, interface strain, and structural modifications could be introduced, thus leading to many extraordinary phenomena. Although this field is still emerging compared to bulk GO composites, it will likely see significant activity in the next decade.

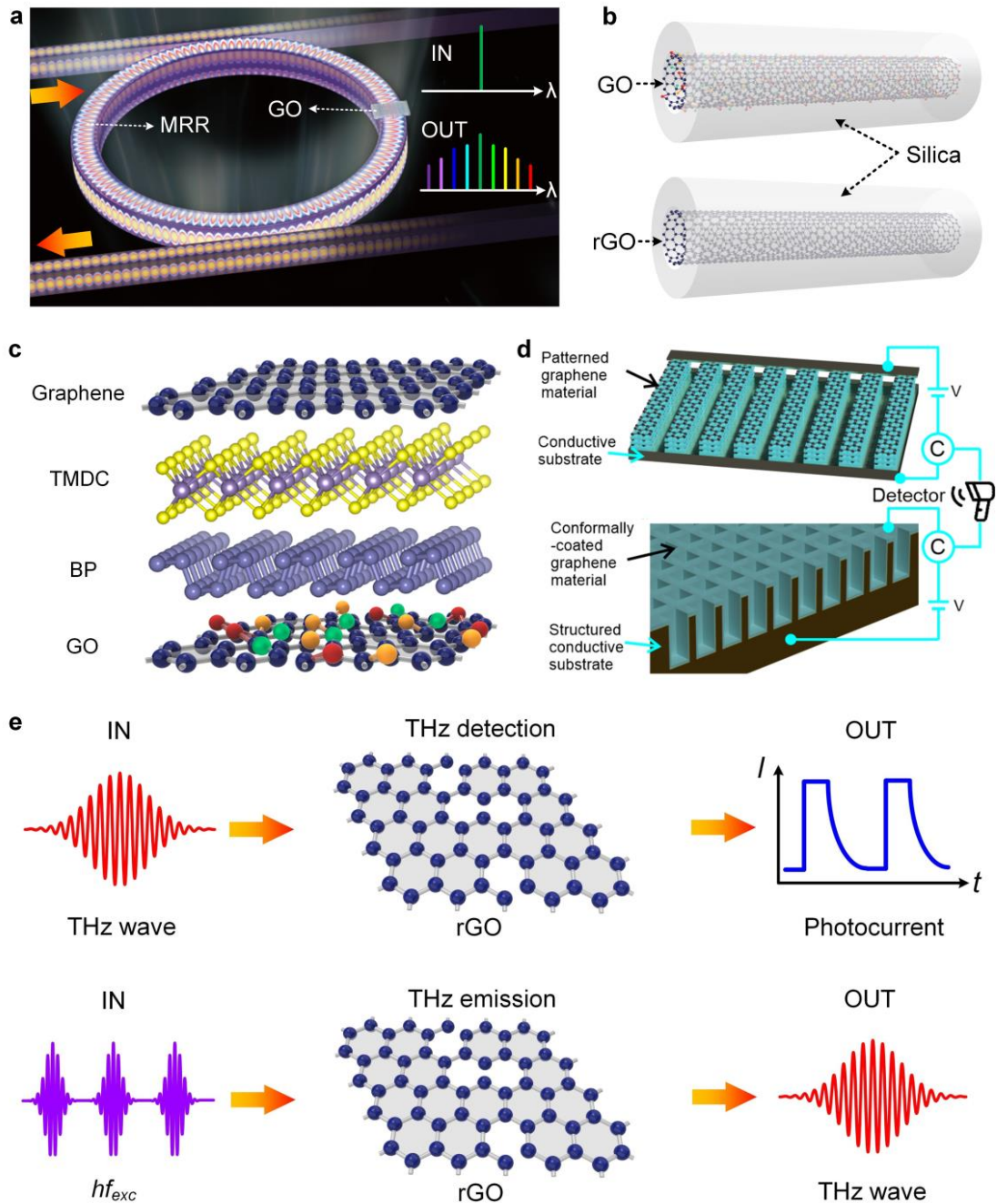

Fig. 10. **Future applications. a|** Optical micro-comb generation and mode locking based on a GO-coated integrated microring resonator (MRR). **b|** Silica optical fibers with embedded GO and rGO on the internal walls. **c|** GO 2D Van der Waals heterostructure. **d** Electrically tunable rGO devices with feedback from a detector. C: control circuitry. V: voltage source. **e|** Terahertz detection and emission based on rGO films.

Due to the high conductivity of rGO, it is possible to dynamically tune its response by applying an external voltage to a conductive coating on a substrate (**FIG. 10d**). In addition, it is possible to use detectors (e.g., PD, thermal camera or optical camera) to monitor the photocurrent output and compare with the designed modulation for calibration. Thus, electrically tunable rGO devices can be realized to optimize the device performance.

Current GO and rGO optoelectronic devices operate mainly at visible and infrared wavelengths. In contrast, the THz regime, which is highly relevant to spectroscopy, imaging, communications, and astronomy[227], remains less explored. In principle, rGO with a bandgap on the order of 10's of meV can be used for THz wave detection. In addition, it can also be used as a source for THz wave emission after absorbing photons with energies larger than its bandgap (**FIG. 10e**). Although many high-performance THz modulators, detectors, and emitters based on graphene have been demonstrated[228-230], the use of rGO for THz applications is still in its infancy[147,231]. A key challenge is to accurately control the material bandgap when it is less than 0.1 eV. This places stringent demands on the material fabrication and characterization methods.

Given the flourish of graphene plasmonics in the past decade[232,233], the use of rGO for plasmonic based technologies is another attractive field, not only from the perspective of the mass production of graphene from GO but also from the new features enabled by localized defects and engineered OFGs.

From an engineering and practicality viewpoint, success in the laboratory at accurately controlling the layer number and position of GO and rGO films has been fueling breakthroughs that will enable mass producible GO films for large scale integration in high-tech industrial sectors, particularly in photonics, electronics, and optoelectronics. The potential of GO films to enhance the performance of nonlinear optical devices such as Kerr microcombs is significant.[234-291] With the continual accumulation of knowledge of GO's material properties, recent progress in device designs,[292-298] continued advances in material synthesis, fabrication, and characterization technologies,[299-300] it is anticipated that major breakthroughs will continue, further bridging the gap between theory and practice and paving the way to deliver the promises of that low-dimensional carbon materials offer.

## Competing interests

The authors declare no competing interests.

# References

1 Kroto, H. W., Heath, J. R., O'Brien, S. C., Curl, R. F. & Smalley, R. E. C60: Buckminsterfullerene. *Nature* **318**, 162-163 (1985).
2 Iijima, S. Helical microtubules of graphitic carbon. *Nature* **354**, 56-58 (1991).
3 Novoselov, K. S. *et al.* Electric field effect in atomically thin carbon films. *Science* **306**, 666-669 (2004).
4 Brodie, B. C. Xiii. On the atomic weight of graphite. *Philosophical Transactions of the Royal Society of London* **149**, 249-259 (1859).
5 Loh, K. P., Bao, Q., Eda, G. & Chhowalla, M. Graphene oxide as a chemically tunable platform for optical applications. *Nat. Chem.* **2**, 1015-1024 (2010).
6 Novoselov, K. S. *et al.* Two-dimensional atomic crystals. *Proc. Natl. Acad. Sci. U. S. A.* **102**, 10451 (2005).
7 Li, L. *et al.* Black phosphorus field-effect transistors. *Nat. Nanotechnol.* **9**, 372-377 (2014).
8 Hummers, W. S. & Offeman, R. E. Preparation of graphitic oxide. *J. Am. Chem. Soc.* **80**, 1339-1339 (1958).
9 Erickson, K. *et al.* Determination of the local chemical structure of graphene oxide and reduced graphene oxide. *Adv. Mater.* **22**, 4467-4472 (2010).
10 Gao, W. *et al.* Direct laser writing of micro-supercapacitors on hydrated graphite oxide films. *Nat. Nanotechnol.* **6**, 496-500 (2011).
11 Voiry, D. *et al.* High-quality graphene via microwave reduction of solution-exfoliated graphene oxide. *Science* **353**, 1413 (2016).
12 Wu, J. *et al.* Graphene oxide waveguide and micro‐ring resonator polarizers. *Laser Photonics Rev.*, 1900056 (2019).
13 Brisebois, P. P. & Siaj, M. Harvesting graphene oxide – years 1859 to 2019: A review of its structure, synthesis, properties and exfoliation. *J. Mater. Chem. C* **8**, 1517-1547 (2020).
14 Dong, L. *et al.* Reactivity-controlled preparation of ultralarge graphene oxide by chemical expansion of graphite. *Chem. Mater.* **29**, 564-572 (2017).
15 Zhang, J. *et al.* Reduction of graphene oxide via l-ascorbic acid. *Chem. Commun. (Camb.)* **46**, 1112-1114 (2010).
16 Dikin, D. A. *et al.* Preparation and characterization of graphene oxide paper. *Nature* **448**, 457-460 (2007).
17 Yang, Y. Y. *et al.* Graphene-based multilayered metamaterials with phototunable architecture for on-chip photonic devices. *ACS Photonics* **6**, 1033-1040 (2019).
18 Zhang, W. H. *et al.* Graphene oxide membranes with stable porous structure for ultrafast water transport. *Nat. Nanotechnol.* (2021).
19 Wu, J. *et al.* Graphene oxide for integrated photonics and flat optics. *Adv. Mater.* **33**, 2006415 (2020).
20 Ghofraniha, N. & Conti, C. Graphene oxide photonics. *J. Opt.* **21**, 053001 (2019).
21 Dideikin, A. T. & Vul, A. Y. Graphene oxide and derivatives: The place in graphene family. *Front. Phys.* **6** (2019).
22 Eigler, S. *et al.* Wet chemical synthesis of graphene. *Adv. Mater.* **25**, 3583-3587 (2013).
23 Yan, J. A., Xian, L. & Chou, M. Y. Structural and electronic properties of oxidized graphene. *Phys. Rev. Lett.* **103**, 086802 (2009).
24 Marcano, D. C. *et al.* Improved synthesis of graphene oxide. *ACS Nano* **4**, 4806-4814 (2010).
25 Pei, S. & Cheng, H.-M. The reduction of graphene oxide. *Carbon* **50**, 3210-3228 (2012).
26 Zheng, X. *et al.* Free-standing graphene oxide mid-infrared polarizers. *Nanoscale* **12**, 11480-11488 (2020).
27 Wong, S. I., Lin, H., Sunarso, J., Wong, B. T. & Jia, B. Triggering a self-sustaining reduction of graphene oxide for high-performance energy storage devices. *ACS Applied Nano Materials* (2020).
28 Lin, K. T., Lin, H., Yang, T. & Jia, B. Structured graphene metamaterial selective absorbers for high efficiency and omnidirectional solar thermal energy conversion. *Nat. Commun.* **11**, 1389 (2020).
29 Hong, J. *et al.* Terahertz conductivity of reduced graphene oxide films. *Optics Express* **21**, 7633-7640 (2013).
30 Furio, A. *et al.* Light irradiation tuning of surface wettability, optical, and electric properties of graphene oxide thin films. *Nanotechnology* **28**, 054003 (2016).
31 Guo, Y. *et al.* General route toward patterning of graphene oxide by a combination of wettability modulation and spin-coating. *ACS Nano* **4**, 5749-5754 (2010).
32 Zheng, X. *et al.* Highly efficient and ultra-broadband graphene oxide ultrathin lenses with three-dimensional subwavelength focusing. *Nat. Commun.* **6**, 8433 (2015).
33 Park, H., Lim, S., Nguyen, D. D. & Suk, J. W. Electrical measurements of thermally reduced graphene oxide powders under pressure. *Nanomaterials* **9**, 1387 (2019).

34  Chua, C. K. & Pumera, M. The reduction of graphene oxide with hydrazine: Elucidating its reductive capability based on a reaction-model approach. *Chemical Communications* **52**, 72-75 (2016).
35  Gusev, A. *et al.* Medium-dependent antibacterial properties and bacterial filtration ability of reduced graphene oxide. *Nanomaterials* **9**, 1454 (2019).
36  Li, X. *et al.* Graphene metalens for particle nanotracking. *Photonics Res.* **8**, 1316 (2020).
37  Zheng, X., Lin, H., Yang, T. & Jia, B. Laser trimming of graphene oxide for functional photonic applications. *Journal of Physics D: Applied Physics* (2016).
38  Yang, T., Lin, H., Zheng, X., Loh, K. P. & Jia, B. Tailoring pores in graphene-based materials: From generation to applications. *Journal of Materials Chemistry A* **5**, 16537-16558 (2017).
39  Zhao, Y., Han, Q., Cheng, Z., Jiang, L. & Qu, L. Integrated graphene systems by laser irradiation for advanced devices. *Nano Today* **12**, 14-30 (2017).
40  Zhang, Y. *et al.* Direct imprinting of microcircuits on graphene oxides film by femtosecond laser reduction. *Nano Today* **5**, 15-20 (2010).
41  Zhang, Y.-L., Chen, Q.-D., Xia, H. & Sun, H.-B. Designable 3d nanofabrication by femtosecond laser direct writing. *Nano Today* **5**, 435-448 (2010).
42  Guo, L. *et al.* Laser‐mediated programmable n doping and simultaneous reduction of graphene oxides. *Advanced Optical Materials* **2**, 120-125 (2014).
43  Zhang, Y. L. *et al.* Photoreduction of graphene oxides: Methods, properties, and applications. *Advanced Optical Materials* **2**, 10-28 (2014).
44  Li, X. H. *et al.* A green chemistry of graphene: Photochemical reduction towards monolayer graphene sheets and the role of water adlayers. *ChemSusChem* **5**, 642-646 (2012).
45  Bagri, A. *et al.* Structural evolution during the reduction of chemically derived graphene oxide. *Nature Chemistry* **2**, 581 (2010).
46  Levis, R. J., Menkir, G. M. & Rabitz, H. Selective bond dissociation and rearrangement with optimally tailored, strong-field laser pulses. *Science* **292**, 709-713 (2001).
47  Williams, G., Seger, B. & Kamat, P. V. Tio2-graphene nanocomposites. Uv-assisted photocatalytic reduction of graphene oxide. *ACS Nano* **2**, 1487-1491 (2008).
48  Huang, L. *et al.* Pulsed laser assisted reduction of graphene oxide. *Carbon* **49**, 2431-2436 (2011).
49  Prezioso, S. *et al.* Large area extreme-uv lithography of graphene oxide via spatially resolved photoreduction. *Langmuir* **28**, 5489-5495 (2012).
50  Chang, H.-W., Tsai, Y.-C., Cheng, C.-W., Lin, C.-Y. & Wu, P.-H. Reduction of graphene oxide in aqueous solution by femtosecond laser and its effect on electroanalysis. *Electrochemistry Communications* **23**, 37-40 (2012).
51  Lin, H. *et al.* A 90-nm-thick graphene metamaterial for strong and extremely broadband absorption of unpolarized light. *Nat. Photonics* **13**, 270-276 (2019).
52  Zheng, X. *et al.* Free-standing graphene oxide mid-infrared polarizers. *Nanoscale* (2020).
53  Wu, J. *et al.* 2d layered graphene oxide films integrated with micro-ring resonators for enhanced nonlinear optics. *Small*, e1906563 (2020).
54  Zhang, L. *et al.* Inkjet printing high-resolution, large-area graphene patterns by coffee-ring lithography. *Adv. Mater.* **24**, 436-440 (2012).
55  Dua, V. *et al.* All-organic vapor sensor using inkjet-printed reduced graphene oxide. *Angew. Chem. Int. Ed. Engl.* **49**, 2154-2157 (2010).
56  Le, L. T., Ervin, M. H., Qiu, H., Fuchs, B. E. & Lee, W. Y. Graphene supercapacitor electrodes fabricated by inkjet printing and thermal reduction of graphene oxide. *Electrochem. Commun.* **13**, 355-358 (2011).
57  Bae, S. *et al.* Roll-to-roll production of 30-inch graphene films for transparent electrodes. *Nature Nanotechnology* **5**, 574-578 (2010).
58  El-Kady, M. F., Strong, V., Dubin, S. & Kaner, R. B. Laser scribing of high-performance and flexible graphene-based electrochemical capacitors. *Science* **335**, 1326 (2012).
59  El-Kady, M. F. & Kaner, R. B. Scalable fabrication of high-power graphene micro-supercapacitors for flexible and on-chip energy storage. *Nat. Commun.* **4**, 1475 (2013).
60  Cao, G. *et al.* Resilient graphene ultrathin flat lens in aerospace, chemical, and biological harsh environments. *ACS Appl. Mater. Interfaces* **11**, 20298-20303 (2019).
61  Wei, S. *et al.* A varifocal graphene metalens for broadband zoom imaging covering the entire visible region. *ACS Nano* (2021).
62  Cao, G., Gan, X., Lin, H. & Jia, B. An accurate design of graphene oxide ultrathin flat lens based on rayleigh-sommerfeld theory. *Opto-Electronic Advances* **1**, 18001201-18001207 (2018).

63. Zhang, H., Yang, D., Lei, C., Lin, H. & Jia, B. Ultrahigh heating rate induced micro-explosive production of graphene for energy storage. *Journal of Power Sources* **442**, 227224 (2019).
64. Gao, M., Zhu, L., Peh, C. K. & Ho, G. W. Solar absorber material and system designs for photothermal water vaporization towards clean water and energy production. *Energy & Environmental Science* **12**, 841-864 (2019).
65. Hu, X. *et al.* Tailoring graphene oxide-based aerogels for efficient solar steam generation under one sun. *Adv. Mater.* **29** (2017).
66. Li, X. *et al.* Graphene oxide-based efficient and scalable solar desalination under one sun with a confined 2d water path. *Proceedings of the National Academy of Sciences* **113**, 13953-13958 (2016).
67. Wong, S. I. *et al.* Tailoring reduction extent of flash-reduced graphene oxides for high performance supercapacitors. *Journal of Power Sources* **478**, 228732 (2020).
68. Wong, S. I., Lin, H., Sunarso, J., Wong, B. T. & Jia, B. Optimization of ionic-liquid based electrolyte concentration for high-energy density graphene supercapacitors. *Applied Materials Today* **18**, 100522 (2020).
69. Wong, S. I. *et al.* Towards enhanced energy density of graphene-based supercapacitors: Current status, approaches, and future directions. *Journal of Power Sources* **396**, 182-206 (2018).
70. Dong, L., Yang, J., Chhowalla, M. & Loh, K. P. Synthesis and reduction of large sized graphene oxide sheets. *Chem. Soc. Rev.* **46**, 7306-7316 (2017).
71. Compton, O. C. & Nguyen, S. T. Graphene oxide, highly reduced graphene oxide, and graphene: Versatile building blocks for carbon-based materials. *Small* **6**, 711-723 (2010).
72. Li, X. *et al.* Athermally photoreduced graphene oxides for three-dimensional holographic images. *Nat. Commun.* **6**, 6984 (2015).
73. Shen, T. Z., Hong, S. H. & Song, J. K. Electro-optical switching of graphene oxide liquid crystals with an extremely large kerr coefficient. *Nat. Mater.* **13**, 394-399 (2014).
74. Hu, Q., Lin, K. T., Lin, H., Zhang, Y. & Jia, B. Graphene metapixels for dynamically switchable structural color. *ACS Nano* **15**, 8930-8939 (2021).
75. Bao, Q. *et al.* Broadband graphene polarizer. *Nat. Photonics* **5**, 411-415 (2011).
76. Lin, H. *et al.* Chalcogenide glass-on-graphene photonics. *Nat. Photonics* **11**, 798-805 (2017).
77. Tan, Y. *et al.* Polarization-dependent optical absorption of mos(2) for refractive index sensing. *Sci. Rep.* **4**, 7523 (2014).
78. Yan, Y. *et al.* High-capacity millimetre-wave communications with orbital angular momentum multiplexing. *Nat. Commun.* **5**, 4876 (2014).
79. Dai, D. X., Bauters, J. & Bowers, J. E. Passive technologies for future large-scale photonic integrated circuits on silicon: Polarization handling, light non-reciprocity and loss reduction. *Light-Science & Applications* **1** (2012).
80. Dai, D. X., Liu, L., Gao, S. M., Xu, D. X. & He, S. L. Polarization management for silicon photonic integrated circuits. *Laser Photonics Rev.* **7**, 303-328 (2013).
81. Lim, W. H. *et al.* Graphene oxide-based waveguide polariser: From thin film to quasi-bulk. *Opt. Express* **22**, 11090-11098 (2014).
82. Chong, W. S. *et al.* Configurable te-and tm-pass graphene oxide-coated waveguide polarizer. *IEEE Photonics Technology Letters*, 1-1 (2020).
83. Guan, X. W. *et al.* Low-loss ultracompact transverse-magnetic-pass polarizer with a silicon subwavelength grating waveguide. *Opt. Lett.* **39**, 4514-4517 (2014).
84. Dai, D. X., Wang, Z., Julian, N. & Bowers, J. E. Compact broadband polarizer based on shallowly-etched silicon-on-insulator ridge optical waveguides. *Opt. Express* **18**, 27404-27415 (2010).
85. Behabtu, N. *et al.* Spontaneous high-concentration dispersions and liquid crystals of graphene. *Nat. Nanotechnol.* **5**, 406-411 (2010).
86. Kim, J. E. *et al.* Graphene oxide liquid crystals. *Angew. Chem. Int. Ed. Engl.* **50**, 3043-3047 (2011).
87. Narayan, R., Kim, J. E., Kim, J. Y., Lee, K. E. & Kim, S. O. Graphene oxide liquid crystals: Discovery, evolution and applications. *Adv. Mater.* **28**, 3045-3068 (2016).
88. Zakri, C. *et al.* Liquid crystals of carbon nanotubes and graphene. *Philos Trans A Math Phys Eng Sci* **371**, 20120499 (2013).
89. Foster, M. A. *et al.* Silicon-chip-based ultrafast optical oscilloscope. *Nature* **456**, 81-84 (2008).
90. Xu, X. *et al.* 11 tops photonic convolutional accelerator for optical neural networks. *Nature* **589**, 44-51 (2021).
91. Jiang, T., Kravtsov, V., Tokman, M., Belyanin, A. & Raschke, M. B. Ultrafast coherent nonlinear nanooptics and nanoimaging of graphene. *Nat. Nanotechnol.* **14**, 838-843 (2019).

92  Yin, X. *et al.* Edge nonlinear optics on a $MoS_2$ atomic monolayer. *Science* **344**, 488 (2014).
93  Roztocki, P. & Morandotti, R. Astrocombs for extreme-precision spectroscopy. *Nat. Astron.* **3**, 135-136 (2019).
94  Li, G., Zentgraf, T. & Zhang, S. Rotational doppler effect in nonlinear optics. *Nat. Phys.* **12**, 736-740 (2016).
95  Kues, M. *et al.* On-chip generation of high-dimensional entangled quantum states and their coherent control. *Nature* **546**, 622-626 (2017).
96  Zhong, H.-S. *et al.* Quantum computational advantage using photons. *Science* **370**, 1460 (2020).
97  Leuthold, J., Koos, C. & Freude, W. Nonlinear silicon photonics. *Nat. Photonics* **4**, 535-544 (2010).
98  Moss, D. J., Morandotti, R., Gaeta, A. L. & Lipson, M. New cmos-compatible platforms based on silicon nitride and hydex for nonlinear optics. *Nat. Photonics* **7**, 597-607 (2013).
99  Autere, A. *et al.* Nonlinear optics with 2d layered materials. *Adv. Mater.* **30**, e1705963 (2018).
100 Mathkar, A. *et al.* Controlled, stepwise reduction and band gap manipulation of graphene oxide. *J Phys Chem Lett* **3**, 986-991 (2012).
101 Yang, Y. *et al.* Invited article: Enhanced four-wave mixing in waveguides integrated with graphene oxide. *APL Photonics* **3**, 120803 (2018).
102 Yoon, H. J. *et al.* Sensitive capture of circulating tumour cells by functionalized graphene oxide nanosheets. *Nat. Nanotechnol.* **8**, 735-741 (2013).
103 Lee, J., Koo, J., Debnath, P., Song, Y. W. & Lee, J. H. A q-switched, mode-locked fiber laser using a graphene oxide-based polarization sensitive saturable absorber. *Laser Phys. Lett.* **10**, 035103 (2013).
104 Ahmad, H., Soltani, S., Thambiratnam, K., Yasin, M. & Tiu, Z. C. Mode-locking in er-doped fiber laser with reduced graphene oxide on a side-polished fiber as saturable absorber. *Opt. Fiber Technol.* **50**, 177-182 (2019).
105 Russier-Antoine, I. *et al.* Second harmonic scattering from mass characterized 2d graphene oxide sheets. *Chem. Commun. (Camb.)* **56**, 3859-3862 (2020).
106 Zhang, Y. *et al.* Enhanced kerr nonlinearity and nonlinear figure of merit in silicon nanowires integrated with 2d graphene oxide films. *ACS Appl. Mater. Interfaces* (2020).
107 Zheng, P. & Wu, N. Fluorescence and sensing applications of graphene oxide and graphene quantum dots: A review. *Chem. Asian J.* **12**, 2343-2353 (2017).
108 Luo, Z., Vora, P. M., Mele, E. J., Johnson, A. T. C. & Kikkawa, J. M. Photoluminescence and band gap modulation in graphene oxide. *Appl. Phys. Lett.* **94**, 111909 (2009).
109 Morales-Narvaez, E. & Merkoci, A. Graphene oxide as an optical biosensing platform. *Adv. Mater.* **24**, 3298-3308 (2012).
110 Cushing, S. K., Li, M., Huang, F. & Wu, N. Origin of strong excitation wavelength dependent fluorescence of graphene oxide. *ACS Nano* **8**, 1002-1013 (2014).
111 Liu, F., Choi, J. Y. & Seo, T. S. Graphene oxide arrays for detecting specific DNA hybridization by fluorescence resonance energy transfer. *Biosens. Bioelectron.* **25**, 2361-2365 (2010).
112 He, S. *et al.* A graphene nanoprobe for rapid, sensitive, and multicolor fluorescent DNA analysis. *Adv. Funct. Mater.* **20**, 453-459 (2010).
113 Liu, X., Guo, Q. & Qiu, J. Emerging low-dimensional materials for nonlinear optics and ultrafast photonics. *Adv. Mater.* **29** (2017).
114 Zhang, B. *et al.* Recent progress in 2d material‐based saturable absorbers for all solid‐state pulsed bulk lasers. *Laser Photonics Rev.* **14**, 1900240 (2019).
115 Wang, G., Baker‐Murray, A. A. & Blau, W. J. Saturable absorption in 2d nanomaterials and related photonic devices. *Laser Photonics Rev.* **13**, 1800282 (2019).
116 Bao, Q. *et al.* Atomic-layer graphene as a saturable absorber for ultrafast pulsed lasers. *Adv. Funct. Mater.* **19**, 3077-3083 (2009).
117 Yu, S. *et al.* All-optical graphene modulator based on optical kerr phase shift. *Optica* **3**, 541 (2016).
118 Li, W. *et al.* Ultrafast all-optical graphene modulator. *Nano Lett.* **14**, 955-959 (2014).
119 Xiaohui, L. *et al.* Broadband saturable absorption of graphene oxide thin film and its application in pulsed fiber lasers. *IEEE J. Sel. Top. Quantum Electron.* **20**, 441-447 (2014).
120 Yasin, M., Thambiratnam, K., Soltani, S. & Ahmad, H. Highly stable mode-locked fiber laser with graphene oxide-coated side-polished d-shaped fiber saturable absorber. *Opt. Eng.* **57**, 1 (2018).
121 Zhao, X. *et al.* Ultrafast carrier dynamics and saturable absorption of solution-processable few-layered graphene oxide. *Appl. Phys. Lett.* **98**, 121905 (2011).
122 Eda, G. *et al.* Blue photoluminescence from chemically derived graphene oxide. *Adv. Mater.* **22**, 505-+ (2010).

123 Zheng, X., Jia, B., Chen, X. & Gu, M. In situ third-order non-linear responses during laser reduction of graphene oxide thin films towards on-chip non-linear photonic devices. *Adv. Mater.* **26**, 2699-2703 (2014).
124 Zhang, X. *et al.* Symmetry-breaking-induced nonlinear optics at a microcavity surface. *Nat. Photonics* **13**, 21-24 (2018).
125 Cazzanelli, M. *et al.* Second-harmonic generation in silicon waveguides strained by silicon nitride. *Nat. Mater.* **11**, 148-154 (2011).
126 Sipe, J. E., Moss, D. J. & van Driel, H. M. Phenomenological theory of optical second- and third-harmonic generation from cubic centrosymmetric crystals. *Phys. Rev. B* **35**, 1129-1141 (1987).
127 DeLong, K. W., Trebino, R., Hunter, J. & White, W. E. Frequency-resolved optical gating with the use of second-harmonic generation. *J. Opt. Soc. Am. B* **11**, 2206-2215 (1994).
128 Wang, C. *et al.* Integrated lithium niobate electro-optic modulators operating at cmos-compatible voltages. *Nature* **562**, 101-+ (2018).
129 Jiang, Y., Tomov, I., Wang, Y. & Chen, Z. Second-harmonic optical coherence tomography. *Opt. Lett.* **29**, 1090-1092 (2004).
130 Furst, J. U. *et al.* Quantum light from a whispering-gallery-mode disk resonator. *Phys. Rev. Lett.* **106**, 113901 (2011).
131 Seyler, K. L. *et al.* Electrical control of second-harmonic generation in a wse2 monolayer transistor. *Nat. Nanotechnol.* **10**, 407-411 (2015).
132 Chen, H. *et al.* Enhanced second-harmonic generation from two-dimensional mose2 on a silicon waveguide. *Light: Sci. Appl.* **6**, e17060 (2017).
133 Fernandes, G. E., Kim, J. H., Osgood, R. & Xu, J. Field-controllable second harmonic generation at a graphene oxide heterointerface. *Nanotechnology* **29**, 105201 (2018).
134 Salem, R. *et al.* Signal regeneration using low-power four-wave mixing on silicon chip. *Nat Photonics* **2**, 35-38 (2008).
135 Ferrera, M. *et al.* Low-power continuous-wave nonlinear optics in doped silica glass integrated waveguide structures. *Nat. Photonics* **2**, 737-740 (2008).
136 Qu, Y. *et al.* Enhanced four‐wave mixing in silicon nitride waveguides integrated with 2d layered graphene oxide films. *Advanced Optical Materials*, 2001048 (2020).
137 Eda, G., Mattevi, C., Yamaguchi, H., Kim, H. & Chhowalla, M. Insulator to semimetal transition in graphene oxide. *The Journal of Physical Chemistry C* **113**, 15768-15771 (2009).
138 Chang, Y.-C. *et al.* Realization of mid-infrared graphene hyperbolic metamaterials. *Nature communications* **7**, 10568 (2016).
139 Bi, Y.-G. *et al.* Arbitrary shape designable microscale organic light-emitting devices by using femtosecond laser reduced graphene oxide as a patterned electrode. *ACS Photonics* **1**, 690-695 (2014).
140 Chen, X., Jia, B., Zhang, Y. & Gu, M. Exceeding the limit of plasmonic light trapping in textured screen-printed solar cells using al nanoparticles and wrinkle-like graphene sheets. *Light: Sci. Appl.* **2**, e92-e92 (2013).
141 Bao, W. *et al.* Flexible, high temperature, planar lighting with large scale printable nanocarbon paper. *Advanced Materials* **28**, 4684-4691 (2016).
142 El-Kady, M. F. & Kaner, R. B. Scalable fabrication of high-power graphene micro-supercapacitors for flexible and on-chip energy storage. *Nature Communications* **4**, 1475 (2013).
143 Tian, H. *et al.* A graphene-based resistive pressure sensor with record-high sensitivity in a wide pressure range. *Scientific Reports* **5**, 8603 (2015).
144 Aspermair, P. *et al.* Reduced graphene oxide–based field effect transistors for the detection of e7 protein of human papillomavirus in saliva. *Analytical and Bioanalytical Chemistry* **413**, 779-787 (2021).
145 Huang, J.-Q. *et al.* Permselective graphene oxide membrane for highly stable and anti-self-discharge lithium–sulfur batteries. *ACS Nano* **9**, 3002-3011 (2015).
146 Wang, X. *et al.* A spectrally tunable all-graphene-based flexible field-effect light-emitting device. *Nat. Commun.* **6**, 7767 (2015).
147 Cao, Y. *et al.* Ultra-broadband photodetector for the visible to terahertz range by self-assembling reduced graphene oxide-silicon nanowire array heterojunctions. *Small* **10**, 2345-2351 (2014).
148 Yeh, T.-F., Syu, J.-M., Cheng, C., Chang, T.-H. & Teng, H. Graphite oxide as a photocatalyst for hydrogen production from water. *Adv. Funct. Mater.* **20**, 2255-2262 (2010).
149 Hsu, H. C. *et al.* Graphene oxide as a promising photocatalyst for co2 to methanol conversion. *Nanoscale* **5**, 262-268 (2013).
150 Han, N. *et al.* Improved heat dissipation in gallium nitride light-emitting diodes with embedded graphene oxide pattern. *Nat. Commun.* **4**, 1452 (2013).

151 Koppens, F. H. *et al.* Photodetectors based on graphene, other two-dimensional materials and hybrid systems. *Nat. Nanotechnol.* **9**, 780-793 (2014).
152 Rogalski, A. Graphene-based materials in the infrared and terahertz detector families: A tutorial. *Advances in Optics and Photonics* **11**, 314 (2019).
153 Mueller, T., Xia, F. & Avouris, P. Graphene photodetectors for high-speed optical communications. *Nat. Photonics* **4**, 297-301 (2010).
154 Gan, X. *et al.* Chip-integrated ultrafast graphene photodetector with high responsivity. *Nat. Photonics* **7**, 883-887 (2013).
155 Wang, X., Cheng, Z., Xu, K., Tsang, H. K. & Xu, J.-B. High-responsivity graphene/silicon-heterostructure waveguide photodetectors. *Nat. Photonics* **7**, 888-891 (2013).
156 Liu, C. H., Chang, Y. C., Norris, T. B. & Zhong, Z. Graphene photodetectors with ultra-broadband and high responsivity at room temperature. *Nat. Nanotechnol.* **9**, 273-278 (2014).
157 Sun, X. *et al.* Broadband photodetection in a microfiber-graphene device. *Opt. Express* **23**, 25209-25216 (2015).
158 Chang, H. *et al.* Thin film field-effect phototransistors from bandgap-tunable, solution-processed, few-layer reduced graphene oxide films. *Adv. Mater.* **22**, 4872-4876 (2010).
159 Ghosh, S., Sarker, B. K., Chunder, A., Zhai, L. & Khondaker, S. I. Position dependent photodetector from large area reduced graphene oxide thin films. *Appl. Phys. Lett.* **96**, 163109 (2010).
160 Lin, Y. *et al.* Dramatically enhanced photoresponse of reduced graphene oxide with linker-free anchored cdse nanoparticles. *ACS Nano* **4**, 3033-3038 (2010).
161 Chitara, B., Krupanidhi, S. B. & Rao, C. N. R. Solution processed reduced graphene oxide ultraviolet detector. *Appl. Phys. Lett.* **99**, 113114 (2011).
162 Chitara, B., Panchakarla, L. S., Krupanidhi, S. B. & Rao, C. N. R. Infrared photodetectors based on reduced graphene oxide and graphene nanoribbons. *Adv. Mater.* **23**, 5419-5424 (2011).
163 Chang, H. *et al.* Regulating infrared photoresponses in reduced graphene oxide phototransistors by defect and atomic structure control. *ACS Nano* **7**, 6310-6320 (2013).
164 Cao, Y., Zhu, J., Xu, J. & He, J. Tunable near-infrared photovoltaic and photoconductive properties of reduced graphene oxide thin films by controlling the number of reduced graphene oxide bilayers. *Carbon* **77**, 1111-1122 (2014).
165 Zhu, M. *et al.* Vertical junction photodetectors based on reduced graphene oxide/silicon schottky diodes. *Nanoscale* **6**, 4909-4914 (2014).
166 Ito, Y. *et al.* 3d bicontinuous nanoporous reduced graphene oxide for highly sensitive photodetectors. *Adv. Funct. Mater.* **26**, 1271-1277 (2016).
167 Li, G. *et al.* Self-powered uv-near infrared photodetector based on reduced graphene oxide/n-si vertical heterojunction. *Small* **12**, 5019-5026 (2016).
168 Moon, I. K., Ki, B., Yoon, S., Choi, J. & Oh, J. Lateral photovoltaic effect in flexible free-standing reduced graphene oxide film for self-powered position-sensitive detection. *Sci. Rep.* **6**, 33525 (2016).
169 Cao, Y. *et al.* Fully suspended reduced graphene oxide photodetector with annealing temperature-dependent broad spectral binary photoresponses. *ACS Photonics* **4**, 2797-2806 (2017).
170 Tian, H., Cao, Y., Sun, J. & He, J. Enhanced broadband photoresponse of substrate-free reduced graphene oxide photodetectors. *RSC Adv.* **7**, 46536-46544 (2017).
171 Yin, Z. *et al.* Graphene-based materials for solar cell applications. *Adv. Energy Mater.* **4**, 1300574 (2014).
172 Liu, Z. *et al.* Organic photovoltaic devices based on a novel acceptor material: Graphene. *Adv. Mater.* **20**, 3924-3930 (2008).
173 Gupta, V. *et al.* Luminscent graphene quantum dots for organic photovoltaic devices. *J. Am. Chem. Soc.* **133**, 9960-9963 (2011).
174 Yang, K. *et al.* Graphene in mice: Ultrahigh in vivo tumor uptake and efficient photothermal therapy. *Nano Lett.* **10**, 3318-3323 (2010).
175 Robinson, J. T. *et al.* Ultrasmall reduced graphene oxide with high near-infrared absorbance for photothermal therapy. *J. Am. Chem. Soc.* **133**, 6825-6831 (2011).
176 Li, M., Yang, X., Ren, J., Qu, K. & Qu, X. Using graphene oxide high near-infrared absorbance for photothermal treatment of alzheimer's disease. *Adv. Mater.* **24**, 1722-1728 (2012).
177 von Maltzahn, G. *et al.* Computationally guided photothermal tumor therapy using long-circulating gold nanorod antennas. *Cancer Res.* **69**, 3892-3900 (2009).
178 Su, Y. *et al.* Impermeable barrier films and protective coatings based on reduced graphene oxide. *Nat. Commun.* **5**, 4843 (2014).

179  Joshi, R. K. *et al.* Precise and ultrafast molecular sieving through graphene oxide membranes. *Science* **343**, 752 (2014).
180  Tu, Y. *et al.* Destructive extraction of phospholipids from escherichia coli membranes by graphene nanosheets. *Nat. Nanotechnol.* **8**, 594-601 (2013).
181  Nine, M. J., Cole, M. A., Tran, D. N. H. & Losic, D. Graphene: A multipurpose material for protective coatings. *J. Mater. Chem. A* **3**, 12580-12602 (2015).
182  Li, H. *et al.* Ultrathin, molecular-sieving graphene oxide membranes for selective hydrogen separation. *Science* **342**, 95 (2013).
183  Shen, J. *et al.* Subnanometer two-dimensional graphene oxide channels for ultrafast gas sieving. *ACS Nano* **10**, 3398-3409 (2016).
184  Qi, B. *et al.* Strict molecular sieving over electrodeposited 2d-interspacing-narrowed graphene oxide membranes. *Nat. Commun.* **8**, 825 (2017).
185  Leo Tsui, H. C. *et al.* Graphene oxide integrated silicon photonics for detection of vapour phase volatile organic compounds. *Sci. Rep.* **10**, 9592 (2020).
186  Chang, D. *et al.* Reversible fusion and fission of graphene oxide–based fibers. *Science* **372**, 614 (2021).
187  Liu, S. *et al.* Antibacterial activity of graphite, graphite oxide, graphene oxide, and reduced graphene oxide: Membrane and oxidative stress. *ACS Nano* **5**, 6971-6980 (2011).
188  Akhavan, O. & Ghaderi, E. Toxicity of graphene and graphene oxide nanowalls against bacteria. *ACS Nano* **4**, 5731-5736 (2010).
189  Guo, F. *et al.* Hydroplastic micromolding of 2d sheets. *Adv. Mater.*, e2008116 (2021).
190  Park, S. *et al.* Graphene oxide papers modified by divalent ions—enhancing mechanical properties via chemical cross-linking. *ACS Nano* **2**, 572-578 (2008).
191  Hu, W. *et al.* Graphene-based antibacterial paper. *ACS Nano* **4**, 4317-4323 (2010).
192  Vallés, C., David Núñez, J., Benito, A. M. & Maser, W. K. Flexible conductive graphene paper obtained by direct and gentle annealing of graphene oxide paper. *Carbon* **50**, 835-844 (2012).
193  Compton, O. C., Dikin, D. A., Putz, K. W., Brinson, L. C. & Nguyen, S. T. Electrically conductive "alkylated" graphene paper via chemical reduction of amine-functionalized graphene oxide paper. *Adv. Mater.* **22**, 892-896 (2010).
194  Chan, K.-Y. *et al.* Graphene oxide thin film structural dielectric capacitors for aviation static electricity harvesting and storage. *Composites Part B: Engineering* **201**, 108375 (2020).
195  Chan, K.-Y., Jia, B., Lin, H., Zhu, B. & Lau, K.-T. Design of a structural power composite using graphene oxide as a dielectric material layer. *Materials Letters* **216**, 162-165 (2018).
196  Li, X., Zhang, Q., Chen, X. & Gu, M. Giant refractive-index modulation by two-photon reduction of fluorescent graphene oxides for multimode optical recording. *Scientific reports* **3** (2013).
197  Iorsh, I. V., Mukhin, I. S., Shadrivov, I. V., Belov, P. A. & Kivshar, Y. S. Hyperbolic metamaterials based on multilayer graphene structures. *Physical Review B* **87** (2013).
198  Koos, C. *et al.* All-optical high-speed signal processing with silicon‑organic hybrid slot waveguides. *Nat. Photonics* **3**, 216-219 (2009).
199  Morelos-Gomez, A. *et al.* Effective nacl and dye rejection of hybrid graphene oxide/graphene layered membranes. *Nat. Nanotechnol.* **12**, 1083-1088 (2017).
200  Lin, H., Jia, B. & Gu, M. Dynamic generation of debye diffraction-limited multifocal arrays for direct laser printing nanofabrication. *Opt. Lett.* **36**, 406-408 (2011).
201  Gómez-Navarro, C. *et al.* Electronic transport properties of individual chemically reduced graphene oxide sheets. *Nano Lett.* **7**, 3499-3503 (2007).
202  Bonaccorso, F., Sun, Z., Hasan, T. & Ferrari, A. C. Graphene photonics and optoelectronics. *Nat. Photonics* **4**, 611-622 (2010).
203  Demetriou, G. *et al.* Nonlinear optical properties of multilayer graphene in the infrared. *Opt. Express* **24**, 13033-13043 (2016).
204  Xu, X. *et al.* Observation of third-order nonlinearities in graphene oxide film at telecommunication wavelengths. *Sci. Rep.* **7**, 9646 (2017).
205  Mu, X., Wu, X., Zhang, T., Go, D. B. & Luo, T. Thermal transport in graphene oxide--from ballistic extreme to amorphous limit. *Sci. Rep.* **4**, 3909 (2014).
206  Balandin, A. A. *et al.* Superior thermal conductivity of single-layer graphene. *Nano Lett.* **8**, 902-907 (2008).
207  Dimiev, A. M., Alemany, L. B. & Tour, J. M. Graphene oxide. Origin of acidity, its instability in water, and a new dynamic structural model. *ACS Nano* **7**, 576-588 (2013).
208  Chien, C. T. *et al.* Tunable photoluminescence from graphene oxide. *Angew. Chem. Int. Ed. Engl.* **51**, 6662-6666 (2012).

209. Galande, C. *et al.* Quasi-molecular fluorescence from graphene oxide. *Sci. Rep.* **1**, 85 (2011).
210. Dong, Y. *et al.* Blue luminescent graphene quantum dots and graphene oxide prepared by tuning the carbonization degree of citric acid. *Carbon* **50**, 4738-4743 (2012).
211. Dave, S. H., Gong, C., Robertson, A. W., Warner, J. H. & Grossman, J. C. Chemistry and structure of graphene oxide via direct imaging. *ACS Nano* **10**, 7515-7522 (2016).
212. Liu, Z. *et al.* Direct observation of oxygen configuration on individual graphene oxide sheets. *Carbon* **127**, 141-148 (2018).
213. Dimiev, A. M. & Polson, T. A. Contesting the two-component structural model of graphene oxide and reexamining the chemistry of graphene oxide in basic media. *Carbon* **93**, 544-554 (2015).
214. Zhang, H. & Miyamoto, Y. Graphene production by laser shot on graphene oxide: Anab initioprediction. *Physical Review B* **85** (2012).
215. Zare, R. N. Laser control of chemical reactions. *Science* **279**, 1875-1879 (1998).
216. Li, D., Mueller, M. B., Gilje, S., Kaner, R. B. & Wallace, G. G. Processable aqueous dispersions of graphene nanosheets. *Nature nanotechnology* **3**, 101-105 (2008).
217. Reed, G. T., Mashanovich, G., Gardes, F. Y. & Thomson, D. J. Silicon optical modulators. *Nat. Photonics* **4**, 518-526 (2010).
218. Reimer, C. *et al.* Generation of multiphoton entangled quantum states by means of integrated frequency combs. *Science* **351**, 1176-1180 (2016).
219. Pasquazi, A. *et al.* Micro-combs: A novel generation of optical sources. *Phys. Rep.* **729**, 1-81 (2018).
220. Spencer, D. T. *et al.* An optical-frequency synthesizer using integrated photonics. *Nature* **557**, 81-85 (2018).
221. Stern, B., Ji, X., Okawachi, Y., Gaeta, A. L. & Lipson, M. Battery-operated integrated frequency comb generator. *Nature* (2018).
222. Yao, B. *et al.* Gate-tunable frequency combs in graphene-nitride microresonators. *Nature* **558**, 410-414 (2018).
223. Agrawal, G. P. Nonlinear fiber optics: Its history and recent progress [invited]. *J. Opt. Soc. Am. B* **28**, A1-A10 (2011).
224. Zuo, Y. *et al.* Optical fibres with embedded two-dimensional materials for ultrahigh nonlinearity. *Nat. Nanotechnol.* **15**, 987-991 (2020).
225. Li, L., Yu, L., Lin, Z. & Yang, G. Reduced tio2-graphene oxide heterostructure as broad spectrum-driven efficient water-splitting photocatalysts. *ACS Appl. Mater. Interfaces* **8**, 8536-8545 (2016).
226. Shi, H. *et al.* A two-dimensional mesoporous polypyrrole-graphene oxide heterostructure as a dual-functional ion redistributor for dendrite-free lithium metal anodes. *Angew. Chem. Int. Ed. Engl.* **59**, 12147-12153 (2020).
227. Ferguson, B. & Zhang, X.-C. Materials for terahertz science and technology. *Nat. Mater.* **1**, 26-33 (2002).
228. Tassin, P., Koschny, T. & Soukoulis, C. M. Graphene for terahertz applications. *Science* **341**, 620 (2013).
229. Vicarelli, L. *et al.* Graphene field-effect transistors as room-temperature terahertz detectors. *Nat. Mater.* **11**, 865-871 (2012).
230. Sensale-Rodriguez, B. *et al.* Broadband graphene terahertz modulators enabled by intraband transitions. *Nat. Commun.* **3**, 780 (2012).
231. Wang, H. *et al.* Terahertz generation from reduced graphene oxide. *Carbon* **134**, 439-447 (2018).
232. Grigorenko, A. N., Polini, M. & Novoselov, K. S. Graphene plasmonics. *Nat. Photonics* **6**, 749-758 (2012).
233. Ju, L. *et al.* Graphene plasmonics for tunable terahertz metamaterials. *Nat. Nanotechnol.* **6**, 630-634 (2011).
234. Xu, X., et al., Photonic microwave true time delays for phased array antennas using a 49 GHz FSR integrated micro-comb source, *Photonics Research*, **6**, B30-B36 (2018).
235. X. Xu, M. Tan, J. Wu, R. Morandotti, A. Mitchell, and D. J. Moss, "Microcomb-based photonic RF signal processing", *IEEE Photonics Technology Letters*, vol. 31 no. 23 1854-1857, 2019.
236. M. Tan et al, "Orthogonally polarized Photonic Radio Frequency single sideband generation with integrated micro-ring resonators", IOP Journal of Semiconductors, Vol. **42** (4), 041305 (2021). DOI: 10.1088/1674-4926/42/4/041305.
237. Xu, *et al.*, "Advanced adaptive photonic RF filters with 80 taps based on an integrated optical micro-comb source," *Journal of Lightwave Technology,* vol. 37, no. 4, pp. 1288-1295 (2019).
238. X. Xu, *et al.*, Broadband microwave frequency conversion based on an integrated optical micro-comb source", *Journal of Lightwave Technology*, vol. 38 no. 2, pp. 332-338 (2020).
239. M. Tan, *et al.*, "Photonic RF and microwave filters based on 49GHz and 200GHz Kerr microcombs", *Optics Comm*. vol. 465,125563 (2020).

240. X. Xu, *et al.,* "Broadband photonic RF channelizer with 90 channels based on a soliton crystal microcomb", *Journal of Lightwave Technology*, Vol. 38, no. 18, pp. 5116 - 5121 (2020). doi: 10.1109/JLT.2020.2997699.
241. X. Xu, *et al.,* "Photonic RF and microwave integrator with soliton crystal microcombs", *IEEE Transactions on Circuits and Systems II: Express Briefs*, vol. 67, no. 12, pp. 3582-3586, 2020. DOI:10.1109/TCSII.2020.2995682.
242. X. Xu, *et al.,* "High performance RF filters via bandwidth scaling with Kerr micro-combs," *APL Photonics,* vol. 4 (2) 026102. 2019.
243. M. Tan, *et al.,* "Microwave and RF photonic fractional Hilbert transformer based on a 50 GHz Kerr micro-comb", *Journal of Lightwave Technology*, vol. 37, no. 24, pp. 6097 – 6104 (2019).
244. M. Tan, *et al.,* "RF and microwave fractional differentiator based on photonics", *IEEE Transactions on Circuits and Systems: Express Briefs*, vol. 67, no.11, pp. 2767-2771 (2020). DOI:10.1109/TCSII.2020.2965158.
245. M. Tan, *et al*., "Photonic RF arbitrary waveform generator based on a soliton crystal micro-comb source", Journal of Lightwave Technology, vol. 38, no. 22, pp. 6221-6226 (2020). DOI: 10.1109/JLT.2020.3009655.
246. M. Tan, X. Xu, J. Wu, R. Morandotti, A. Mitchell, and D. J. Moss, "RF and microwave high bandwidth signal processing based on Kerr Micro-combs", Advances in Physics X, VOL. 6, NO. 1, 1838946 (2021). DOI:10.1080/23746149.2020.1838946.
247. X. Xu, et al., "Advanced RF and microwave functions based on an integrated optical frequency comb source," Opt. Express, vol. 26 (3) 2569 (2018).
248. M. Tan, X. Xu, J. Wu, B. Corcoran, A. Boes, T. G. Nguyen, S. T. Chu, B. E. Little, R.Morandotti, A. Lowery, A. Mitchell, and D. J. Moss, ""Highly Versatile Broadband RF Photonic Fractional Hilbert Transformer Based on a Kerr Soliton Crystal Microcomb", Journal of Lightwave Technology vol. 39 (24) 7581-7587 (2021).
249. Wu, J. *et al.* RF Photonics: An Optical Microcombs' Perspective. IEEE Journal of Selected Topics in Quantum Electronics Vol. **24**, 6101020, 1-20 (2018).
250. T. G. Nguyen *et al.*, "Integrated frequency comb source-based Hilbert transformer for wideband microwave photonic phase analysis," *Opt. Express,* vol. 23, no. 17, pp. 22087-22097 (2015).
251. X. Xu, J. Wu, M. Shoeiby, T. G. Nguyen, S. T. Chu, B. E. Little, R. Morandotti, A. Mitchell, and D. J. Moss, "Reconfigurable broadband microwave photonic intensity differentiator based on an integrated optical frequency comb source," *APL Photonics*, vol. 2, no. 9, 096104 (2017).
252. X. Xu, *et al.*, "Broadband RF channelizer based on an integrated optical frequency Kerr comb source," *Journal of Lightwave Technology,* vol. 36, no. 19, pp. 4519-4526 (2018).
253. X. Xu, *et al.*, "Continuously tunable orthogonally polarized RF optical single sideband generator based on micro-ring resonators," *Journal of Optics,* vol. 20, no. 11, 115701 (2018).
254. X. Xu, *et al.*, "Orthogonally polarized RF optical single sideband generation and dual-channel equalization based on an integrated microring resonator," *Journal of Lightwave Technology,* vol. 36, no. 20, pp. 4808-4818 (2018).
255. M.Tan, X. Xu, J. Wu, B. Corcoran, A. Boes, T. G. Nguyen, Sai T. Chu, B. E. Little, R.Morandotti, A. Mitchell, and D. J. Moss, "Integral order photonic RF signal processors based on a soliton crystal micro-comb source", IOP Journal of Optics vol. 23 (11) 125701 (2021).
256. X. Xu, *et al.,* "Photonic RF phase-encoded signal generation with a microcomb source", *J. Lightwave Technology*, vol. 38, no. 7, 1722-1727 (2020).
257. B. Corcoran, et al., "Ultra-dense optical data transmission over standard fiber with a single chip source", Nature Communications, vol. 11, Article:2568 (2020).
258. X. Xu et al, "Photonic perceptron based on a Kerr microcomb for scalable high speed optical neural networks", Laser and Photonics Reviews, vol. 14, no. 8, 2000070 (2020). DOI: 10.1002/lpor.202000070.
259. X. Xu, et al., "11 TOPs photonic convolutional accelerator for optical neural networks", Nature **589**, 44-51 (2021).
260. Yang Sun, Jiayang Wu, Mengxi Tan, Xingyuan Xu, Yang Li, Roberto Morandotti, Arnan Mitchell, and David Moss, "Applications of optical micro-combs", Advances in Optics and Photonics Vol. 15, Issue 1, pp. 86-175 (2023). DOI: 10.1364/AOP.470264.
261. Yunping Bai, Xingyuan Xu,1, Mengxi Tan, Yang Sun, Yang Li, Jiayang Wu, Roberto Morandotti, Arnan Mitchell, Kun Xu, and David J. Moss, "Photonic multiplexing techniques for neuromorphic computing", Nanophotonics vol. **11** (2022). DOI:10.1515/nanoph-2022-0485.
262. Xingyuan Xu, Weiwei Han, Mengxi Tan, Yang Sun, Yang Li, Jiayang Wu, Roberto Morandotti, Arnan Mitchell, Kun Xu, and David J. Moss, "Neuromorphic computing based on wavelength-division multiplexing", IEEE Journal of Selected Topics in Quantum Electronics vol. 29 Issue: 2, Article 7400112, March-April (2023). DOI:10.1109/JSTQE.2022.3203159.

263. Chawaphon Prayoonyong, Andreas Boes, Xingyuan Xu, Mengxi Tan, Sai T. Chu, Brent E. Little, Roberto Morandotti, Arnan Mitchell, David J. Moss, and Bill Corcoran, "Frequency comb distillation for optical superchannel transmission", Journal of Lightwave Technology vol. 39 (23) 7383-7392 (2021).
264. Mengxi Tan, X. Xu, J. Wu, T. G. Nguyen, S. T. Chu, B. E. Little, R. Morandotti, A. Mitchell, and David J. Moss, "Photonic Radio Frequency Channelizers based on Kerr Optical Micro-combs", IOP Journal of Semiconductors vol. 42 (4), 041302 (2021).
265. Mengxi Tan, Yang Li, Xingyuan Xu, Bill Corcoran, Jiayang Wu, Andreas Boes, Thach G. Nguyen, Sai T. Chu, Brent E. Little, Damien G. Hicks, Roberto Morandotti, Arnan Mitchell, and David J. Moss, "11 Tera-OPs/s photonic convolutional accelerator and deep optical neural network based on an integrated Kerr soliton crystal microcomb", Paper No. LA203-15, SPIE-LASE, Laser Resonators, Microresonators, and Beam Control XXIV, SPIE-Opto, SPIE Photonics West, San Francisco CA January 22 - 27 (2022). doi: 10.1117/12.2607906
266. Mengxi Tan, Yang Sun, Yang Li, Xingyuan Xu, Andreas Boes, Bill Corcoran, Jiayang Wu, Thach G. Nguyen, Sai T. Chu, Brent E. Little, Roberto Morandotti, Arnan Mitchell, and David Moss, "RF and microwave photonic signal generation and processing based on Kerr micro-combs", Paper No. PW22O-OE106-17, SPIE-Opto, Terahertz, RF, Millimeter, and Submillimeter-Wave Technology and Applications XV, SPIE Photonics West, San Francisco CA January 22 - 27 (2022). DOI: 10.1117/12.2607905.
267. Mengxi Tan, Yang Li, Yang Sun, Xingyuan Xu, Jiayang Wu, Sai T. Chu, Brent E. Little, Roberto Morandotti, Arnan Mitchell, and David J. Moss, "Versatile, high bandwidth, RF and microwave photonic Hilbert transformers based on Kerr micro-combs", Paper No. PW22O-OE201-21, SPIE-Opto, Integrated Optics: Devices, Materials, and Technologies XXVI, SPIE Photonics West, San Francisco CA January 22 - 27 (2022). doi: 10.1117/12.2607903
268. Mengxi Tan, Xingyuan Xu, Bill Corcoran, J. Wu, A. Boes, T. G. Nguyen, Sai T. Chu, Brent E. Little, Damien G. Hicks, Roberto Morandotti, Arnan Mitchell, and David J. Moss, "Optical Neuromorphic Processor at 11 TeraOPs/s based on Kerr Soliton Crystal Micro-combs", paper Tu3G.1, OSA/IEEE Optical Fibre Communications Conference, San Diego CA, March 6 - 10 (2022). DOI: 10.1364/OFC.2022.Tu3G.1.
269. Antonio Cutrona, Maxwell Rowley, Debayan Das, Luana Olivieri, Luke Peters, Sai T. Chu, Brent L. Little, Roberto Morandotti, David J. Moss, Juan Sebastian Totero Gongora, Marco Peccianti, Alessia Pasquazi, "High Conversion Efficiency in Laser Cavity-Soliton Microcombs", Optics Express Vol. 30, Issue 22, pp. 39816-39825 (2022).
270. Maxwell Rowley, Pierre-Henry Hanzard, Antonio Cutrona, Hualong Bao, Sai T. Chu, Brent E. Little, Roberto Morandotti, David J. Moss, Gian-Luca Oppo, Juan Sebastian Totero Gongora, Marco Peccianti and Alessia Pasquazi, "Self-emergence of robust solitons in a micro-cavity", Nature **608,** (7922) 303 – 309 (2022).
271. A. Pasquazi, et al., "Sub-picosecond phase-sensitive optical pulse characterization on a chip", Nature Photonics, vol. 5, no. 10, pp. 618-623 (2011).
272. Bao, C., et al., Direct soliton generation in microresonators, Opt. Lett, **42**, 2519 (2017).
273. M.Ferrera et al., "CMOS compatible integrated all-optical RF spectrum analyzer", Optics Express, vol. 22, no. 18, 21488 - 21498 (2014).
274. M. Kues, et al., "Passively modelocked laser with an ultra-narrow spectral width", Nature Photonics, vol. 11, no. 3, pp. 159, 2017.
275. L. Razzari, et al., "CMOS-compatible integrated optical hyper-parametric oscillator," Nature Photonics, vol. 4, no. 1, pp. 41-45, 2010.
276. M. Ferrera, et al., "Low-power continuous-wave nonlinear optics in doped silica glass integrated waveguide structures," Nature Photonics, vol. 2, no. 12, pp. 737-740, 2008.
277. M.Ferrera et al."On-Chip ultra-fast 1st and 2nd order CMOS compatible all-optical integration", Opt. Express, vol. 19, (23)pp. 23153-23161 (2011).
278. D. Duchesne, M. Peccianti, M. R. E. Lamont, et al., "Supercontinuum generation in a high index doped silica glass spiral waveguide," Optics Express, vol. 18, no, 2, pp. 923-930, 2010.
279. H Bao, L Olivieri, M Rowley, ST Chu, BE Little, R Morandotti, DJ Moss, ... "Turing patterns in a fiber laser with a nested microresonator: Robust and controllable microcomb generation", Physical Review Research **2** (2), 023395 (2020).
280. M. Ferrera, et al., "On-chip CMOS-compatible all-optical integrator", Nature Communications, vol. 1, Article 29, 2010.
281. A. Pasquazi, et al., "All-optical wavelength conversion in an integrated ring resonator," Optics Express, vol. 18, no. 4, pp. 3858-3863, 2010.
282. A.Pasquazi, Y. Park, J. Azana, et al., "Efficient wavelength conversion and net parametric gain via Four Wave Mixing in a high index doped silica waveguide," Optics Express, vol. 18, no. 8, pp. 7634-7641, 2010.

283. M. Peccianti, M. Ferrera, L. Razzari, et al., "Subpicosecond optical pulse compression via an integrated nonlinear chirper," Optics Express, vol. 18, no. 8, pp. 7625-7633, 2010.
284. Little, B. E. et al., "Very high-order microring resonator filters for WDM applications", IEEE Photonics Technol. Lett. **16**, 2263–2265 (2004).
285. M. Ferrera et al., "Low Power CW Parametric Mixing in a Low Dispersion High Index Doped Silica Glass Micro-Ring Resonator with Q-factor > 1 Million", Optics Express, vol.17, no. 16, pp. 14098–14103 (2009).
286. M. Peccianti, et al., "Demonstration of an ultrafast nonlinear microcavity modelocked laser", Nature Communications, vol. 3, pp. 765, 2012.
287. A.Pasquazi, et al., "Self-locked optical parametric oscillation in a CMOS compatible microring resonator: a route to robust optical frequency comb generation on a chip," Optics Express, vol. 21, no. 11, pp. 13333-13341, 2013.
288. A.Pasquazi, et al., "Stable, dual mode, high repetition rate mode-locked laser based on a microring resonator," Optics Express, vol. 20, no. 24, pp. 27355-27362, 2012.
289. Pasquazi, A. et al. Micro-combs: a novel generation of optical sources. Physics Reports **729**, 1-81 (2018).
290. Moss, D. J. et al., "New CMOS-compatible platforms based on silicon nitride and Hydex for nonlinear optics", Nature photonics vol. **7**, 597 (2013).
291. H. Bao, et al., Laser cavity-soliton microcombs, Nature Photonics, vol. 13, no. 6, pp. 384-389, Jun. 2019.
292. Yuning Zhang, Jiayang Wu, Linnan Jia, Yang Qu, Baohua Jia, and David J. Moss, "Graphene oxide for nonlinear integrated photonics", Laser and Photonics Reviews vol. 16 (2023). DOI:10.1002/lpor.202200512
293. Yang Qu, Jiayang Wu, Yuning Zhang, Yunyi Yang, Linnan Jia, Baohua Jia, and David J. Moss, "Photo thermal tuning in GO-coated integrated waveguides", Micromachines vol. 13 1194 (2022). doi.org/10.3390/mi13081194
294. Yuning Zhang, Jiayang Wu, Yunyi Yang, Yang Qu, Houssein El Dirani, Romain Crochemore, Corrado Sciancalepore, Pierre Demongodin, Christian Grillet, Christelle Monat, Baohua Jia, and David J. Moss, "Enhanced self-phase modulation in silicon nitride waveguides integrated with 2D graphene oxide films", IEEE Journal of Selected Topics in Quantum vol. 28 (2022). DOI: 10.1109/JSTQE.2022.3177385
295. Yuning Zhang, Jiayang Wu, Yunyi Yang, Yang Qu, Linnan Jia, Baohua Jia, and David J. Moss, "Enhanced spectral broadening of femtosecond optical pulses in silicon nanowires integrated with 2D graphene oxide films", Micromachines vol. 13 756 (2022).
296. Yuning Zhang, Jiayang Wu, Yang Qu, Linnan Jia, Baohua Jia, and David J. Moss, "Optimizing the Kerr nonlinear optical performance of silicon waveguides integrated with 2D graphene oxide films", Journal of Lightwave Technology vol. 39 (14) 4671-4683 (2021).
297. Yang Qu, Jiayang Wu, Yuning Zhang, Yao Liang, Baohua Jia, and David J. Moss, "Analysis of four-wave mixing in silicon nitride waveguides integrated with 2D layered graphene oxide films", Journal of Lightwave Technology vol. 39 (9) 2902-2910 (2021).
298. Yuning Zhang, Jiayang Wu, Yang Qu, Yang Sun, Yunyi Yang, Linnan Jia, Baohua Jia, and David Moss, "Enhanced nonlinear optics in nanowires, waveguides, and ring resonators integrated with graphene oxide films", Paper No. PW22O-OE108-11, Oxide-based Materials and Devices XIII, SPIE Photonics West, San Francisco CA March 6-11 (2022). DOI:10.1117/12.2607904
299. Linnan Jia, Jiayang Wu, Yuning Zhang, Yang Qu, Baohua Jia, Zhigang Chen, and David J. Moss, "Fabrication Technologies for the On-Chip Integration of 2D Materials", Small: Methods vol. 6 2101435 (2022). DOI:10.1002/smtd.202101435.
300. Linnan Jia, Changfu Huo, Xiaoqing Yan, Jiayang Wu, Yunyi Yang, Daohong Song, Baohua Jia, Zhibo Liu, Zhigang Chen, David J. Moss, "Ultrafast carrier dynamics in 2D NbTe2 films", ACS Applied Nano Materials, vol. **5** (12), 17348–17355 (2022). DOI: 10.1021/acsanm.2c04333.